\documentclass[sigconf]{acmart}

\AtBeginDocument{%
  \providecommand\BibTeX{{%
    \normalfont B\kern-0.5em{\scshape i\kern-0.25em b}\kern-0.8em\TeX}}}

\usepackage{flushend}
\usepackage{caption}
\usepackage{tabularx}
\usepackage{graphicx}
\usepackage{subfigure}
\usepackage{amsfonts}
\usepackage{amsmath}
\usepackage{amsthm}
\usepackage{algorithm}
\usepackage{algorithmic}
\usepackage{multirow}
\usepackage{color}
\usepackage{xcolor}
\usepackage{booktabs}
\usepackage{float}
\usepackage{balance}
\usepackage{xspace}
\usepackage{threeparttable}
\usepackage{url}

\newcommand{\ie}{\textit{i}.\textit{e}.}
\newcommand{\eg}{\textit{e}.\textit{g}.} 
\newcommand{\etal}{\textit{et al}.}
\newcommand{\wrt}{\textit{w}.\textit{r}.\textit{t}}
\newtheorem{Def}{Definition}
\newtheorem*{Problem}{Problem}

\newcommand{\system}{MHGCN\xspace}
\newcommand{\varr}{MHGCN-R\xspace}
\newcommand{\varl}{MHGCN-L\xspace}
\def\model{MHGCN}

\copyrightyear{2022}
\acmYear{2022}
\setcopyright{acmcopyright}\acmConference[KDD '22]{Proceedings of the 28th ACM SIGKDD Conference on Knowledge Discovery and Data Mining}{August 14--18, 2022}{Washington, DC, USA}
\acmBooktitle{Proceedings of the 28th ACM SIGKDD Conference on Knowledge Discovery and Data Mining (KDD'22), August 14--18, 2022, Washington, DC, USA}
\acmPrice{15.00}
\acmDOI{10.1145/3534678.3539482}
\acmISBN{978-1-4503-9385-0/22/08}

\begin{document}

\title{Multiplex Heterogeneous Graph Convolutional Network}



\author{Pengyang Yu}
\affiliation{
  \institution{Ocean University of China}
  \city{Qingdao}
  \country{China}
  \postcode{266100}
}
\email{ypy@stu.ouc.edu.cn}

\author{Chaofan Fu}
\affiliation{%
  \institution{Ocean University of China}
  \city{Qingdao}
  \country{China}}
\email{fuchaofan@stu.ouc.edu.cn}

\author{Yanwei Yu}
\authornote{Yanwei Yu and Chao Huang are the corresponding authors.}
\affiliation{%
  \institution{Ocean University of China}
  \city{Qingdao}
  \country{China}}
\email{yuyanwei@ouc.edu.cn}

\author{Chao Huang}
\authornotemark[1]
\affiliation{
  \institution{The University of Hong Kong}
  \city{Hong Kong}
  \country{China}}
\email{chaohuang75@gmail.com}

\author{Zhongying Zhao}
\affiliation{
  \institution{Shandong University of Science and Technology}
  \city{Qingdao}
  \country{China}
}
\email{zyzhao@sdust.edu.cn}

\author{Junyu Dong}
\affiliation{%
  \institution{Ocean University of China}
  \city{Qingdao}
  \country{China}
}
\email{dongjunyu@ouc.edu.cn}

\renewcommand{\shortauthors}{Pengyang Yu et al.}

\begin{abstract}

Heterogeneous graph convolutional networks have gained great popularity in tackling various network analytical tasks on heterogeneous network data, ranging from link prediction to node classification. However, most existing works ignore the relation heterogeneity with multiplex network between multi-typed nodes and different importance of relations in meta-paths for node embedding, which can hardly capture the heterogeneous structure signals across different relations. 
To tackle this challenge, this work proposes a \underline{\textbf{M}}ultiplex \underline{\textbf{H}}eterogeneous  \underline{\textbf{G}}raph  \underline{\textbf{C}}onvolutional  \underline{\textbf{N}}etwork (\system) for heterogeneous network embedding. 
Our \system can automatically learn the useful heterogeneous meta-path interactions of different lengths in multiplex heterogeneous networks through multi-layer convolution aggregation. Additionally, we effectively integrate both multi-relation structural signals and attribute semantics into the learned node embeddings with both unsupervised and semi-supervised learning paradigms.  
Extensive experiments on five real-world datasets with various network analytical tasks demonstrate the significant superiority of \system against state-of-the-art embedding baselines in terms of all evaluation metrics. The source code of our method is available at: \url{https://github.com/NSSSJSS/MHGCN}.

\end{abstract}




\maketitle

\section{Introduction}

Network representation learning has emerged as a new learning paradigm to embed complex network into a low-dimensional vector space while preserving the proximities of nodes in both network topological structures and intrinsic properties. Effective network representation advances various network analytical tasks, ranging from link prediction~\cite{DBLP:conf/kdd14Deepwalk,chen2018pme,liu2021motif}, node classification~\cite{wu2019simplifying,liu2020fast,grover2016node2vec}, to recommendation~\cite{shi2018heterogeneous,ji2021heterogeneous, 2021knowledge}. In recent years, Graph Convolutional Networks (GCNs)~\cite{iclr17KipfGCN}, a class of neural networks designed to learn graph representation for complex networks with rich feature information, have been applied to many online services, such as E-commerce~\cite{li2020hierarchical}, social media platforms~\cite{wu2020graph} and advertising~\cite{he2021click}.  


While many efforts have been made to study the representation learning over homogeneous graphs~\cite{tang2015line,grover2016node2vec,qiu2019netsmf,iclr17KipfGCN}, the exploration of preserving network heterogeneous properties in graph representation paradigms has attracted much attention in recent studies, \eg, metapath2vec~\cite{dong2017metapath2vec} and HERec~\cite{shi2018heterogeneous}. Inspired by the strength of Graph Neural Networks (GNNs) in aggregating contextual signals from neighboring nodes, various graph neural models have been introduced to tackle the challenge of heterogeneous graph learning, such as HAN~\cite{www2019HAN}, MAGNN~\cite{fu2020magnn} and HetGNN~\cite{zhang2019heterogeneous}.

Albeit the effectiveness of existing heterogeneous network embedding methods~\cite{ji2021heterogeneous,dong2017metapath2vec,lu2019relation,wang2020dynamic}, these works are generally designed for heterogeneous networks with a single view. In real-world scenarios, however, many networks are much more complex, comprising not only multi-typed nodes and diverse edges even between the same pair-wise nodes but also a rich set of attributes~\cite{kdd19GATNE}. For example, in E-commerce networks, there are two types of nodes (\ie, users and items), and multiple relations (\eg, click, purchase, add-to-cart, or add-to-preference) between the same pairs of users and items~\cite{2020multiplex}. The connections between multiple types of nodes in such networks are often heterogeneous with relation diversity, which yields networks with multiple different views. It is worth noting that the multiplicity of the network is fundamentally different from the heterogeneity of the network. 
Two types of nodes, users and items, in a E-commerce network reflect the heterogeneity of the network. 
At the same time, users may have several types of interactions (\eg, click, purchse, review) with items~\cite{wei2022contrastive}, which reflects the multiplex relationships of the network. Because different user-item interactions exhibit different views of user and item, and thus should be treated differently. We term this kind of networks with both multiplex network structures with multi-typed nodes and node attribute information as \textbf{\textit{attributed multiplex heterogeneous networks}} (AMHENs). 

Performing representation learning on the AMHENs is of great importance to network mining tasks, yet it is very challenging due to such complicated network structures and node attributes. While some recent studies propose to solve the representation learning problem on multiplex heterogeneous network~\cite{kdd19GATNE,nips19GTN,park2019DMGI,liu2020fast,xue2021multiplex}, several key limitations exist in those methods. i) The success of current representation learning models largely relies on the accurate design of meta-paths. How to design an automated learning framework to explore the complex meta-path-based dependencies over the multiplex heterogeneous graphs, remains a significant challenge. ii) Unlike the homogeneous node aggregation scheme, with the heterogeneous node types and multiplex node relationships, each meta-path can be regarded as relational information channel. An effective meta-path dependency encoder is a necessity to inject both the relation heterogeneity and multiplexity into the representations. iii) In real-world graph representation scenarios, efficiency is an important factor to handle the graph data with large number of heterogeneous nodes and multiplex edges. However, most current methods are limited to serve the large-scale network data, due to their high time complexity and memory consumption.

To address the aforementioned challenges, we propose a new \underline{\textbf{M}}ultiplex  \underline{\textbf{H}}eterogeneous  \underline{\textbf{G}}raph \underline{\textbf{C}}onvolutional  \underline{\textbf{N}}etwork, named \textbf{\system}, for AMHEN embedding. 
Specifically, we first decouple the multiplex network into multiple homogeneous and bipartite sub-networks, and then re-aggregate the sub-networks with the exploration of their importance (\ie, weights) in node representation learning. 
To automatically capture meta-path information across multi-relations, we tactfully design a multilayer graph convolution module, which can effectively learn the useful heterogeneous meta-path interactions of different lengths in AMHENs through multilayer convolution aggregation in both unsupervised and semi-supervised learning paradigms. To improve the model efficiency, we endow our \system with a simplified graph convolution for feature aggregation, in order to significantly reduce the model computational cost. 
Our evaluations are conducted on several real-world graph datasets to evaluate the model performance in both link prediction and node classification tasks. Experimental results show that our \system framework can obtain the substantial performance improvement compared with state-of-the-art graph representation techniques. With the designed graph convolution module, our \system achieves better model efficiency when competing with state-of-the-art GNN baselines for AMHENs by up to two orders of magnitudes (see efficiency analysis in the supplemental material). .



We summarize the contributions of this paper as follows:
\begin{itemize}

\item We propose an effective multiplex heterogeneous graph neural network, \system, which can automatically capture the useful relation-aware topological structural signals between nodes for heterogeneous network embedding. \\\vspace{-0.05in}

\item \system integrates both network structures and node attribute features in node representations, and gains the capability to efficiently learn network representation with a simplified convolution-based message passing mechanism. \\\vspace{-0.05in}

\item We conduct extensive experiments on five real-world datasets to verify the superiority of our proposed model in both link prediction and node classification when competing with state-of-the-art baselines.  

\end{itemize}

\begin{figure*}
    \begin{center}
    \vspace{-0mm}
    \includegraphics[width=0.99\textwidth]{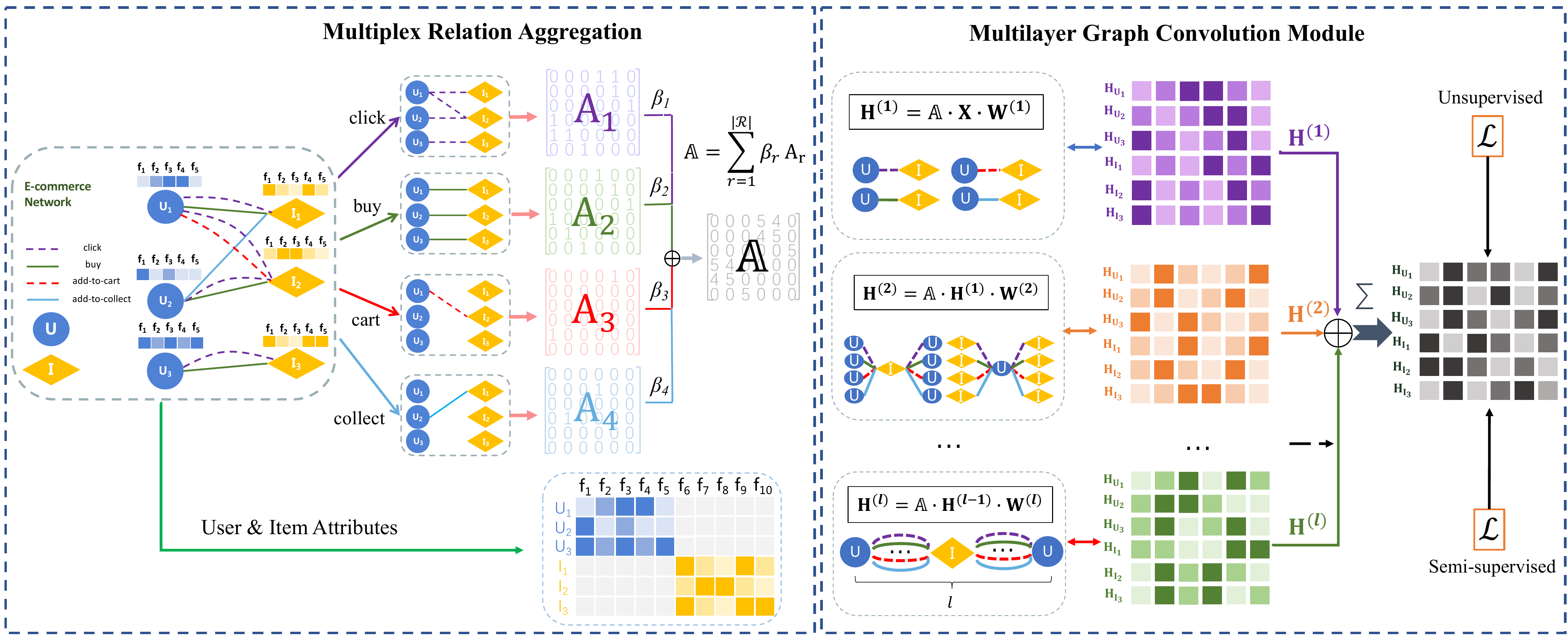}
    \vspace{-2mm}
    \caption{The overview of the proposed \system.}
    \label{fig:framework} 
    \vspace{-2mm}
    \end{center}
\end{figure*}

\section{Related Work}

\textbf{Graph Neural Networks.} 
The goal of a GNN is to learn a low-dimensional vector representation for each node, which can be used for many downstream network mining tasks. Kipf~\etal~\cite{iclr17KipfGCN} proposes to perform convolutional operations over graph neighboring node for information aggregation.
GraphSAGE~\cite{William2017GraphSAGE} is an inductive GNN framework, which uses the general aggregating functions for efficient generation of node embeddings. 
To differentiate the influence of neighboring nodes, GAT~\cite{DBLP:conf/iclr18GAT} has been proposed as an attentive message passing mechanism to learn the explicit weights of neighbor node embeddings. R-GCN~\cite{schlichtkrull2018modeling} considers the influence of different edge types on nodes, and uses weight sharing and coefficient constraints to apply to multi-graphs with large numbers of relations. To simplify the design of graph convolutional network, LightGCN~\cite{he2020lightgcn} omits the embedding projection with non-linearity during the message passing. Additionally, AM-GCN~\cite{wang2020gcn} is proposed to adaptively learn deep correlation information between topological structures and node features. However, all algorithms mentioned above are developed for the homogeneous networks, and thus cannot effectively preserve the heterogeneous and multiplex graph characteristics for the network representation task. \\\vspace{-0.1in}



\noindent \textbf{Heterogeneous Graph Representation.}
Modeling the heterogeneous context of graphs has already received some attention~\cite{dong2017metapath2vec,shi2018heterogeneous,zhang2019heterogeneous,2021recent,long2021social}. For example, some studies leverage random walks to construct meta-paths over the heterogeneous graph for node embeddings, including metapath2vec~\cite{dong2017metapath2vec} and HERec~\cite{shi2018heterogeneous}. As graph neural networks (GNNs) have become a popular choice for encoding graph structures, many heterogeneous graph neural network models are designed to enhance the GNN architecture with the capability of capturing the node and edge heterogeneous contextual signals. For example, HetGNN~\cite{zhang2019heterogeneous} jointly encodes the graph topology and context heterogeneity for representation learning. 
HeGAN~\cite{kdd19HeGAN} incorporates generative adversarial networks (GAN) for heterogeneous network embedding. NARS~\cite{yu2020scalable} first generates relation subgraphs, learns node embeddings by 1D convolution on the subgraphs and then aggregates the learned embeddings. Fu~\etal~\cite{fu2020magnn} performs both the intra- and inter-metapath aggregation so as to distill the metapath-based relational context for learning node representations. However, most of those approaches rely on selecting the useful metapaths to guide the process of heterogeneous representation, which may need the external domain knowledge for constructing relevant metapaths.

In addition, there exist some recent studies attempting to relax the requirement of metapath construction for heterogeneous graph representations. In particular, HGT~\cite{hu2020heterogeneous} proposes to incorporate the self-attention into the graph-based message passing mechanism for modeling the dynamic dependencies among heterogeneous nodes. HPN~\cite{ji2021heterogeneous} eliminates semantic confusion by mapping nodes in meta-path to semantic space, and then aggregates the embeddings of nodes under different metapaths to obtain the final representation. 
However, most of the above heterogeneous graph embedding models ignore the multiplex relational context of real-life graph data, in which multi-typed relationships exist among nodes. 
\noindent \textbf{Multiplex Heterogeneous Network Embedding.}  
Real-world graphs are often inherently multiplex, which involves various relations and interactions between two connected nodes. To tackle this challenge, many multiplex network embedding techniques are proposed to project diverse node edges into latent representations. For example, MNE~\cite{zhang2018scalable} introduces a global transformation matrix for each layer of the network to align the embeddings with different dimensions for each relation type. GATNE~\cite{kdd19GATNE} splits the node representation by learning base embedding, edge embedding as well as attribute embedding. The self-attention is utilized to fuse neighborhood information for generating edge representation. Motivated by the mutual information maximization scheme, DMGI~\cite{park2019DMGI} is proposed as an unsupervised learning approach which aims to minimize the difference among relation-aware node representations. HGSL~\cite{zhao2021heterogeneous} first obtains the node representation based on meta-paths, and then uses GNN to jointly train the heterogeneous graph, node representation and node attributes to obtain the final embedding. However, the generality of the above methods is limited by their manual construction of meta-paths. 

Recently, FAME~\cite{liu2020fast} develops a spectral graph transformation component to aggregate information from sub-networks by preserving relation-aware node dependencies. However, this model is built on the random projection and sacrifices the adaptive parameter learning in exchange for fast embedding projection. Furthermore, to learn the node embeddings of multiplex bipartite graph, DualHGCN~\cite{xue2021multiplex} firstly generates two sets of homogeneous hypergraphs and then perform the information propagation with the spectral hypergraph convolutions. In HDI~\cite{jing2021hdmi}, Jing~\etal~explores the high-order mutual information to construct the supervision signals for enhancing the node representations.



\vspace{-0mm}
\section{Problem Definition}
\label{sec.prob}


We define graph $\mathcal{G}=\{\mathcal{V}, \mathcal{E}\}$ with the set of nodes $\mathcal{V}$ and edges $\mathcal{E}$. Each edge in $\mathcal{E}$ represents the connections among nodes.




\begin{Def}[Attributed Multiplex Heterogeneous Network, or AMHEN]
Given the defined graph $\mathcal{G}$, we further associate all nodes in $\mathcal{V}$ with the attribute feature vectors $\mathbf{X} \in \mathbb{R}^{n\times m}$. Here, the size of node set $\mathcal{V}$ and attribute vector is represented by $n$ and $m$, respectively. With the consideration of node and edge heterogeneity, we define the node type and edge type mapping function as $\phi : \mathcal{V} \to \mathcal{O}$ and $\psi : \mathcal{E} \to \mathcal{R}$. Here, the set of node types and edge types is set with the size of $\mathcal{O}$ and $\mathcal{R}$, respectively. Each node $v\in \mathcal{V}$ and edge $e\in \mathcal{E}$ are associated with a specific type in $\mathcal{O}$ and $\mathcal{R}$, respectively. Additionally, with the consideration of edge multiplexity (\ie, $|\mathcal{O}| + |\mathcal{R}| > 2$), the same pair of nodes can be connected through multi-typed edges.
\end{Def} 

\begin{Def}[Meta-path]
A meta-path $\mathcal{P}$ is defined as a path in the form of $O_1 \xrightarrow{r_1} O_2 \xrightarrow{r_2} \cdots \xrightarrow{r_{l-1}} O_l$ which describes a composite relation $R = r_1 \circ r_2 \cdots r_{l-1}$ between node types $O_1$ and $O_l$, where $\circ$ denotes the composition operator on relations. 
\end{Def} 



For example, $U_1 \xrightarrow{click} I_2 \xrightarrow{buy} U_2$ is a meta-path sample of meta-path $User \xrightarrow{click} Item \xrightarrow{buy} User$. Based on the above definitions, we formally present the representation learning task over the multiplex heterogeneous graph as follows:
\begin{Problem}[Attributed Multiplex Heterogeneous Graph Representation]
The objective of our representation learning task over the attributed multiplex heterogeneous graph $\mathcal{G}=\{ \mathcal{V}, \mathcal{E}, \mathbf{X}\}$ is to learn low-dimensional latent embedding (with the hidden dimensionality of $d$ $d\ll|\mathcal{V}|$) for each node $v\in \mathcal{V}$, with the preservation of node and edge heterogeneity and multiplexity.
\end{Problem}

We summarize the key notations of our technical solution in Table~\ref{table_notations} presented in the supplementary material.


\section{Methodology}
This section describe our framework \system with the overall architecture shown in Figure~\ref{fig:framework}. Particularly, our \system contains two key learning modules: (i) \textit{multiplex relation aggregation} and (ii) \textit{multilayer graph convolution module}. \textit{Multiplex relation aggregation} aims to aggregate the multi-relations among heterogeneous nodes in multiplex heterogeneous networks by differentiating each relation with importance. 
\textit{Multilayer graph convolution module} can automatically capture the heterogeneous meta-paths of different lengths across multi-relations by aggregating neighboring nodes' characteristics to learn the low-dimensional representation of nodes.


\begin{figure*}
    \begin{center}
    \vspace{-0mm}
    \includegraphics[width=0.83\textwidth]{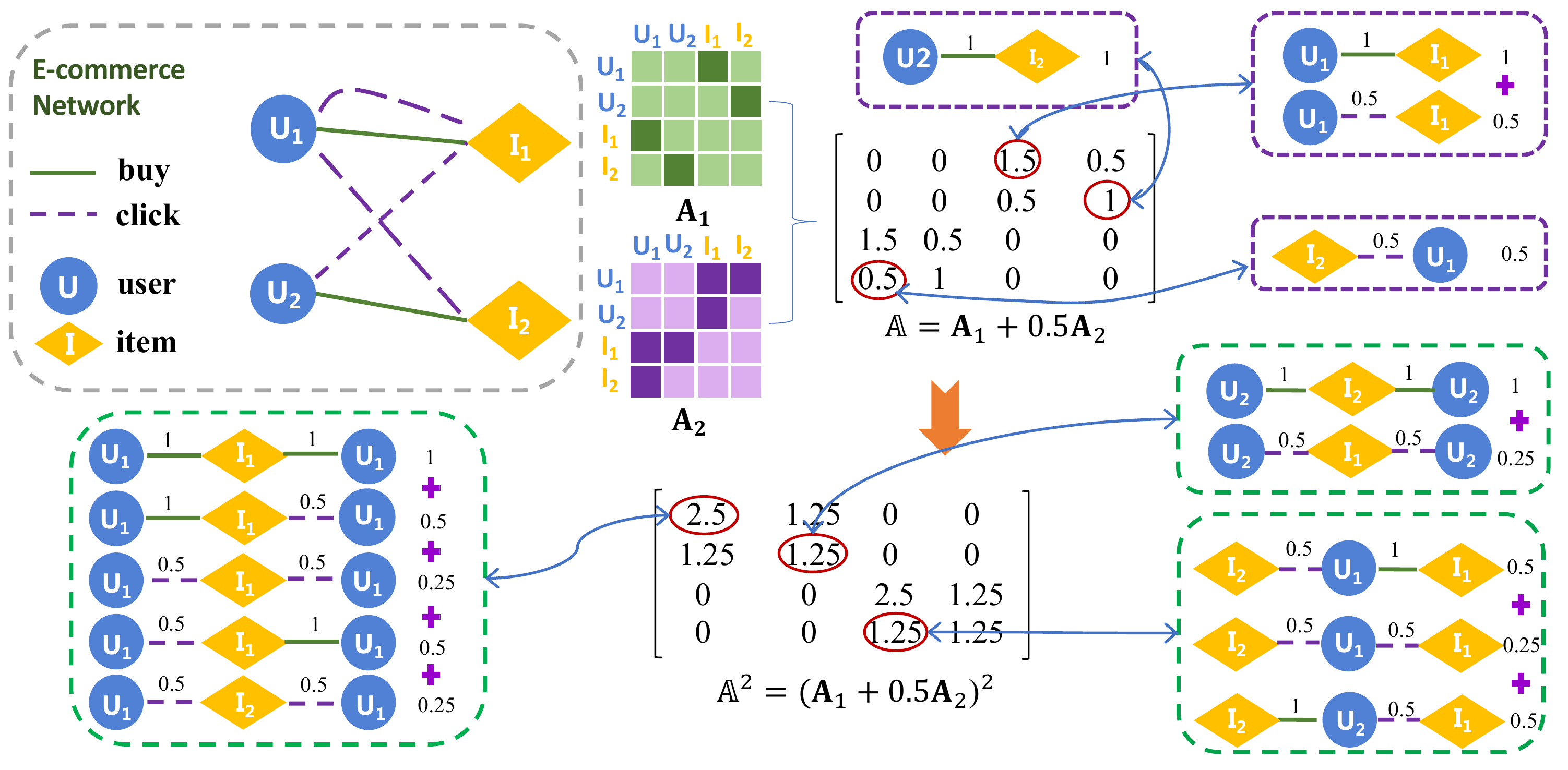}
    \vspace{-2mm}
    \caption{Illustration of meta-paths with importance for a toy example}
    \label{fig:example}   
    \vspace{-4mm}
    \end{center}
\end{figure*}

\subsection{Multiplex Relation Aggregation}

As defined in Sec.~\ref{sec.prob}, there exit different types of nodes and multiple types of edges between these nodes in AMHENs, and each type of edge has a different role and impact on node representation. Therefore, following~\cite{liu2020fast}, we first generate multiple sub-graphs by differentiating the types of edge connections between nodes in the multiplex and heterogeneous graph. Afterwards, we aggregate the relation-aware graph contextual information with different importance weights.



We denote our generated sub-graph as $\{\mathcal{G}_r|r=1,2,\dots,|\mathcal{R}|\}$ with the corresponding adjacent matrix $\{\mathbf{A}_r|r = 1,2,\dots,|\mathcal{R}|\}$. Considering the scenario of multiplex user-item relations in online retailer (\eg, click, purchase, review), the decomposed sub-graph corresponds to individual type of relationship between user and item.
For instance, for the graph representation learning in E-commerce platforms, different relationships (different edge types) between user and item nodes exhibit various dependency semantics. For example, the diverse behaviors of users (\eg, click, add-to-favorite, purchase) reflect different preferences of users over items. Hence, multiplex user-item interactions with various relation semantics will have different impacts on the learning process of user representations. To capture such multi-typed node dependencies, our proposed \system learns the relation-aware weights $\beta_r$ to aggregate edge-type-specific sub-graph adjacent matrix as: $\mathbb{A} = \sum_{r=1}^{|\mathcal{R}|} \beta_r \mathbf{A}_r$.
Notice that the set of weights $\{\beta_r|r=1,2,\dots,|\mathcal{R}|\}$ should not be a set of hyperparameter, but should be dynamically changed according to different tasks, so we set them as trainable parameters to be learned in model training.

\subsection{Multilayer Graph Convolution Module}
Different from homogeneous networks, heterogeneous networks contain different types of nodes and edges. The specified types of edges and nodes form a meta-path, which has an obvious effect on the representation learning of heterogeneous networks. Previous works require manually defined meta-paths and learn node representations on the sampled heterogeneous meta-paths. However, setting and sampling meta-paths artificially is a complex task. In a large-scale network, the number of meta-paths is very large. It takes a long time to sample such a large number of meta-paths. At the same time, aggregating meta-paths into meta-path graph also requires a lot of memory overhead. 
Additionally, the type of meta-paths has an important impact on node representation, which almost determines the performance of network embedding in various downstream tasks. The number of types of heterogeneous meta-paths is also very large, involving different lengths and different relation interactions. Therefore, it is difficult to select the appropriate meta-path types for heterogeneous network embedding methods based on meta-path aggregation. 
Our \system effectively solves the above problems. We now present our multilayer graph convolution module that automatically captures the the short and long meta-paths across multi-relations in AMHENs.


It is worth noting that our model uses a multi-layer fusion GCN. As shown in Figure~\ref{fig:framework}, our graph convolution module consists of multiple graph convolutional layers. Its purpose is to capture meta-path information of different lengths. 
Next, we take a two-layer GCN as an example to illustrate how our model capture meta-path information. For a single layer GCN:
\begin{equation}
\label{Single layer GCN}
    \mathbf{H}^{(1)} = \mathbb{A} \cdot \mathbf{X} \cdot \mathbf{W}^{(1)},
\end{equation}
where $\mathbf{H}^{(1)}\in \mathbb{R}^{n\times d}$ is the output of first layer (\ie, hidden representation of network), $\mathbf{X}\in \mathbb{R}^{n\times m}$ is the node attribute matrix, and $\mathbf{W}^{(1)}\in \mathbb{R}^{m\times d}$ is the learnable weight matrix. 
Notice that our convolution adopts the idea of simplifying GCN~\cite{wu2019simplifying}, that is, no non-linear activation function is used. 

For the two-layer GCN, the message passing process can be represented as below:
\begin{equation}
\label{Double layer GCN}
\begin{split}
    \mathbf{H}^{(2)} &= \mathbb{A} \cdot \mathbf{H}^{(1)} \cdot \mathbf{W}^{(2)}\\
    &= \mathbb{A} \cdot (\mathbb{A} \cdot \mathbf{X} \cdot \mathbf{W}^{(1)}) \cdot \mathbf{W}^{(2)}\\
    &= \mathbb{A}^{2} \cdot \mathbf{X} \cdot \mathbf{W}^{(1)}\cdot \mathbf{W}^{(2)},
\end{split}
\end{equation}
where $\mathbf{W}^{(2)}\in \mathbb{R}^{d\times d}$ is the learnable weight matrix for second layer.  


We present an illustrated example with a graph generated from E-commerce data in Figure~\ref{fig:example} based on two types of node relations, namely users' buy and click behaviors on items. As shown in Figure~\ref{fig:example}, aggregated matrix $\mathbb{A}$ can be regarded as a meta-path graph matrix generated by the 1-length meta-paths with importance (\ie, all linked node pairs across all edge types with weights). For example, $\mathbb{A}_{(1,3)}=1.5$ contains two 1-length meta-path samples with weights, \ie, $U_1 \xrightarrow{1*buy} I_1: 1$ and $U_1 \xrightarrow{0.5*click} I_1: 0.5$. Therefore, the single-layer GCN can effectively learn the node representation that contains 1-length meta-path information. 
Similarly, the second power of $\mathbb{A}$ automatically captures the 2-length meta-path information with importance weights for all node pairs, including original sub-network high-order structures. For example, $\mathbb{A}^2_{(1,1)}=2.5$ implies five 2-length meta-path samples across multi-relations with importance, \ie, $U_1 \xrightarrow{1*buy} I_1 \xrightarrow{1*buy} U_1: 1$, $U_1 \xrightarrow{1*buy} I_1 \xrightarrow{0.5*click} U_1: 0.5$, $U_1 \xrightarrow{0.5*click} I_1 \xrightarrow{0.5*click} U_1: 0.25$, $U_1 \xrightarrow{0.5*click} I_1 \xrightarrow{1*buy} U_1: 0.5$, and $U_1 \xrightarrow{0.5*click} I_2 \xrightarrow{0.5*click} U_1: 0.25$. The sum of the importance of these five meta-path samples is 2.5. 

At the same time, considering that the influence of meta-paths with different lengths on embedding should also be different, the learnable weight matrices $\mathbf{W}^{(l)}$ in our multilayer graph convolution module can just play this role. 
Eventually, we fuse the outputs of single-layer GCN and two-layer GCN:
\begin{equation}
\label{Double layer fusion GCN}
    \mathbf{H} = \frac{1}{2}(\mathbf{H}^{(1)} + \mathbf{H}^{(2)}). 
\end{equation}
The final embedding $\mathbf{H}\in \mathbb{R}^{n\times d}$ contains all 1-length and 2-length meta-path information. 

To capture the more length heterogeneous meta-paths, we can extend it to $l$-layer: 
\begin{equation}
\label{l-layer GCN}
\begin{split}
    \mathbf{H}^{(l)} &= \mathbb{A} \cdot \mathbf{H}^{(l-1)} \cdot \mathbf{W}^{(l)}\\
    &= \mathbb{A} \cdot (\mathbb{A} \cdot \mathbf{H}^{(l-2)} \cdot \mathbf{W}^{(l-1)}) \cdot \mathbf{W}^{(l)}\\
    &= \underbrace{\mathbb{A} \cdots (\mathbb{A}}_l \cdot \mathbf{X} \cdot \underbrace{\mathbf{W}^{(1)}) \cdots \mathbf{W}^{(l)}}_l\\
    &= \mathbb{A}^{l} \cdot \mathbf{X} \cdot \underbrace{\mathbf{W}^{(1)} \cdots \mathbf{W}^{(l)}}_l 
\end{split}
\end{equation}

Therefore, our multilayer graph convolution module fuses outputs of all layers to capture all meta-path information of different length across multi-relations: 
\begin{equation}
\label{Multilayer fusion GCN}
\begin{split}
    \mathbf{H} &= \frac{1}{l}\sum_{i=1}^{l}\mathbf{H}^{(i)}\\
    &=\frac{1}{l} \sum_{i=1}^{l}\mathbb{A} \cdot \mathbf{H}^{(i-1)} \cdot \mathbf{W}^{(i)},
\end{split}
\end{equation}
where $\mathbf{H}^{(0)}$ is the node attribute matrix $\mathbf{X}$.


\vspace{-0mm}
\subsection{Model Learning}


This part presents the defined objective function to train our model to learn the final node representation. Depending on the requirements of different downstream tasks and the availability of node labels, we can train \system in two major learning paradigms, \ie, unsupervised learning and semi-supervised learning.

For unsupervised learning, we can optimize the model parameters by minimizing the following binary cross-entropy loss function through negative sampling: 
\begin{equation}
\label{Unsupervised Loss}
    \begin{split}
     \mathcal{L} &=-\sum_{(u, v) \in \Omega} \log \sigma(<\mathbf{H}^\mathsf{T}_u, \mathbf{H}_v>) - \sum_{(u', v') \in \Omega^{-}}\log\sigma(-<\mathbf{H}^\mathsf{T}_{u'}, \mathbf{H}_{v'}>),
    \end{split}
\end{equation}
where $\mathbf{H}_v$ is the representation of node $v$, $\mathsf{T}$ denotes matrix transposition,  $\sigma(\cdot)$ is the sigmoid function, $<,>$ can be any vector similarity measure function (\eg, inner product), $\Omega$ is the set of positive node pairs, $\Omega^-$ is the set of negative node pairs sampled from all unobserved node pairs. 
That is, we use the loss function to increase the similarities between the node representations in the positive samples and decrease the similarities between the node representations in the negative samples simultaneously. 

For semi-supervised learning, we can optimize the model parameters by minimizing the cross entropy via backpropagation and gradient descent. 
The cross entropy loss over all labeled nodes between the ground-truth and the prediction is formulated as: 
\begin{equation}
\label{Supervised Loss}
     \mathcal{L} = -\sum_{i \in \mathcal{V}_{ids}} \mathbf{Y}_i\ \mathrm{ln}(\mathbf{C} \cdot \mathbf{H}_i), 
\end{equation}
where $\mathcal{V}_{ids}$ is the set of node indices that have labels, $\mathbf{Y}_i$ is the label of the $i$-th node, $\mathbf{C}$ is the node classifier parameter, and $\mathbf{H}_i$ is the representation of the $i$-th node.  
With the guide of a small fraction of labeled nodes, we can optimize the proposed model and then learn the embeddings of nodes for semi-supervised classification. 

Notice that $\{\mathbf{W}^{(i)}|i=1,2,\dots, l\}$ and  $\{\beta_r|r=1,2,\dots,|\mathcal{R}|\}$ in our model can be learned during training phase. 
The pseudo-code of our proposed \system is shown in Algorithm~\ref{alg.FastLANE} in the supplement.

\vspace{-0mm}
\section{Experiment}

\begin{table}[t]
\begin{center}
\vspace{-0mm}
\caption{Statistical information of evaluation network datasets (node type: n-type, edge type: e-type, features: feat., and Multiplex network: Mult.)}
\label{table_dataset} 
\vspace{-2mm}
\setlength{\tabcolsep}{1.5mm}{}
\begin{tabular}{c|c|c|c|c|c|c}
\toprule
Dataset & \#nodes & \#edges & \#n-type & \#e-type & \#feat. & Mult. \\
\midrule
Alibaba & 21,318 & 41,676 & 2 & 4 & 19 & \checkmark \\
Amazon & 10,166 & 148,865 & 1 & 2 & 1,156 & \checkmark \\
AMiner & 58,068 & 118,939 & 3 & 3 & 4 & $\times$ \\
IMDB & 12,772 & 18,644 & 3 & 2 & 1,256 & $\times$ \\
DBLP & 26,128 & 119,783 & 4 & 3 & 4,635 & $\times$ \\
\bottomrule
\end{tabular}    
\end{center}
\vspace{-4mm}
\end{table} 



\subsection{Datasets}

In our evaluation, five publicly available real-world datasets are used in experimental evaluation, \ie, Alibaba\footnote{\url{https://tianchi.aliyun.com/competition/entrance/231719/information/}}, Amazon\footnote{\url{http://jmcauley.ucsd.edu/data/amazon/}}, AMiner\footnote{\url{https://github.com/librahu/}}, IMDB\footnote{\url{https://github.com/seongjunyun/Graph_Transformer_Networks}}, and DBLP\footnote{\url{https://www.dropbox.com/s/yh4grpeks87ugr2/DBLP_processed.zip?dl=0}}. 
Detailed dataset description can be found in the supplement. 
Due to the scalability limitation of applying some baselines in the whole Alibaba networ data, we evaluate all models on a sampled dataset from Alibaba. 
The statistics of these five datasets are summarized in Table~\ref{table_dataset}.  

\begin{table*}[t]
\begin{center}
\vspace{-0mm}
\caption{Model performance comparison for the task of link prediction on different datasets.}
\label{table_linkprediction}
\vspace{-2mm}
\setlength{\tabcolsep}{0.8mm}{}
\begin{threeparttable}
\begin{tabular}{c|ccc|ccc|ccc|ccc|ccc}
\toprule
\multirow{2}{*}{Method} & \multicolumn{3}{c|}{AMiner}  & \multicolumn{3}{c|}{Alibaba} & \multicolumn{3}{c|}{IMDB} & \multicolumn{3}{c|}{Amazon} & \multicolumn{3}{c}{DBLP} \\ 
 & R-AUC & PR-AUC & F1 & R-AUC & PR-AUC & F1 & R-AUC & PR-AUC & F1 & R-AUC & PR-AUC & F1 & R-AUC & PR-AUC & F1 \\ 
\midrule
node2vec & 0.594 & 0.663 & 0.602  & 0.614 & 0.580 & 0.593 & 0.479 & 0.568 & 0.474 & 0.946 & 0.944 & 0.880 & 0.449 & 0.452 & 0.478 \\
RandNE & 0.607 & 0.630 & 0.608  & 0.877  & 0.888 & 0.826 & 0.901 & 0.933 & 0.839 & 0.950 & 0.941 & 0.903 & 0.492 & 0.491 & 0.493 \\
FastRP & 0.620 & 0.634 & 0.600 & 0.927 & 0.900 & 0.926 & 0.869 & 0.893 & 0.811  & 0.954 & 0.945 & 0.893 & 0.515 & 0.528 & 0.506 \\
SGC & 0.589 &	0.585 &	0.567 & 0.686  & 0.708  & 0.623 & 0.826 & 0.889 & 0.769   & 0.791 & 0.802 & 0.760 & 0.601 & 0.606 &	0.587  \\
R-GCN & 0.599 & 0.601 & 0.610 & 0.674 & 0.710 & 0.629 & 0.826 & 0.878 & 0.790 & 0.811 & 0.820 & 0.783 & 0.589 & 0.592 & 0.566 \\
MAGNN & 0.663 & 0.681 & 0.666 & 0.961 & 0.963 & 0.948 & 0.912 & 0.923 & 0.887 & 0.958 & 0.949 & 0.915 & 0.690 & 0.699 & 0.684 \\
HPN & 0.658 & 0.664 & 0.660 & 0.958 & 0.961 & 0.950 & 0.900 & 0.903 & 0.892 & 0.949 & 0.949 & 0.904 & 0.692 & 0.710 & 0.687 \\\hline
PMNE-n & 0.651 & 0.669 & 0.677 & 0.966 & 0.973 & 0.891 & 0.674 & 0.683 & 0.646 & 0.956 & 0.945 & 0.893 & 0.672 & 0.679 & 0.663 \\
PMNE-r & 0.615 & 0.653 & 0.662 & 0.859 & 0.915 & 0.824 & 0.646 & 0.646 & 0.613 & 0.884 & 0.890 & 0.796 & 0.637 & 0.640 & 0.629 \\
PMNE-c & 0.613 & 0.635 & 0.657 & 0.597 & 0.591 & 0.664 & 0.651 & 0.634 & 0.630 & 0.934 & 0.934 & 0.868 & 0.622 & 0.625 & 0.609 \\
MNE & 0.660 & 0.672 & 0.681 & 0.944 & 0.946 & 0.901 & 0.688 & 0.701 & 0.681 & 0.941 & 0.943 & 0.912 & 0.657 & 0.660 & 0.635 \\
GATNE & OOT & OOT & OOT & 0.981 & 0.986 & 0.952 & 0.872 & 0.878 & 0.791 & 0.963 & 0.948 & 0.914 & OOT & OOT & OOT \\
DMGI & OOM & OOM & OOM & 0.857 & 0.781 & 0.784 & 0.926 & 0.935 & 0.873 & 0.905 & 0.878 & 0.847 & 0.610 & 0.615 & 0.601 \\
FAME & 0.687 & 0.747 & 0.726 & 0.993 & 0.996 & 0.979 & 0.944 & 0.959 & 0.897 & 0.959 & 0.950 & 0.900 & 0.642 & 0.650 & 0.633 \\
DualHGNN & / & / & / & 0.974 & 0.977 & 0.966 & / & / & / & / & / & / & / & / & / \\
\hline
\textbf{\system} & \textbf{0.711} & \textbf{0.753} & \textbf{0.730} & \textbf{0.997} & \textbf{0.997} & \textbf{0.992} & \textbf{0.967} & \textbf{0.966} & \textbf{0.959} & \textbf{0.972} & \textbf{0.974} & \textbf{0.961} & \textbf{0.718} & \textbf{0.722} & \textbf{0.703}\\ 
\bottomrule
\end{tabular}
\begin{tablenotes}
        \item OOT: Out Of Time (36 hours). OOM: Out Of Memory; DMGI runs out of memory on the entire AMiner data. R-AUC: ROC-AUC.
\end{tablenotes}
\end{threeparttable}    
\end{center}
\vspace{-3mm}
\end{table*}

\vspace{-0mm}
\subsection{Baselines}
We compare our \system against the following eighteen graph learning baselines, which are divided into three categories.

Homogeneous network embedding methods:
\begin{itemize}


\item \textbf{node2vec}~\cite{grover2016node2vec} - node2vec is a representative method for graph representation by leveraging the random walk to generate node sequences over graphs.


\item \textbf{RandNE}~\cite{icdm18billionNE} - RandNE performs projection process in an iterative way to capture the high-order graph structures with the matrix factorization objective.


\item \textbf{FastRP}~\cite{DBLP:conf/cikm19_fastRP} - This method generates similarity matrix for modeling the transitive relations among nodes. Then, FastRP leverages sparse random projection to reduce dimension.


\item \textbf{SGC}~\cite{wu2019simplifying} - SGC proposes to simplify the graph convolutional networks by removing the non-linear projection during the information propagation between graph layers.

\item \textbf{AM-GCN}~\cite{wang2020gcn} - AM-GNN is a state-of-the-art graph convolutional network, which is an adaptive multi-channel graph convolutional networks for semi-supervised classification.

\end{itemize}

Heterogeneous network embedding methods:
\begin{itemize}
\item \textbf{R-GCN}~\cite{schlichtkrull2018modeling} - R-GCN further considers the influence of different edge types on nodes, and uses weight sharing and coefficient constraints to apply to heterogeneous networks. 

\item \textbf{HAN}~\cite{www2019HAN} - HAN applies graph attention network on multiplex network considering the inter- and intra-network interactions, which exploit manually selected meta-paths to learn node embedding. 

\item \textbf{NARS}~\cite{yu2020scalable} NARS decouples heterogeneous networks according to the type of edge, and then aggregates neighbor features on the decoupled subgraph. 


\item \textbf{MAGNN}~\cite{fu2020magnn} - MAGNN is a metapath aggregated graph neural network for heterogeneous graphs. 

\item \textbf{HPN}~\cite{ji2021heterogeneous} - HPN designs a semantic propagation mechanism to alleviate semantic confusion and a semantic fusion mechanism to integrate rich semantics. 
\end{itemize}

Multiplex Heterogeneous network embedding methods:
\begin{itemize}

\item \textbf{PMNE}~\cite{liu2017PMNE} - PMNE contains three different models to merge the multiplex network to generate one overall embedding for each node, which are denoted as PMNE-n, PMNE-r, and PMNE-c, respectively. 


\item \textbf{MNE}~\cite{zhang2018scalable} - MNE obtains the final embedding by combining the high-dimensional common embedding and the low-dimensional hierarchical embedding. 


\item \textbf{GATNE}~\cite{kdd19GATNE} - GATNE proposes to generate the overall node embeddings with the base embedding as well as the edge and attribute representations. The edge embedding is generated by fusing neighboring information using self-attention.



\item \textbf{GTN}~\cite{nips19GTN} - It studies the graph representation task by identifying effective meta-paths with high-order relations. 

\item \textbf{DMGI}~\cite{park2019DMGI} - DMGI 
develops a consensus regularization scheme to consider the relations among type-specific node embeddings. Furthermore, each relational contextual signals are aggregated through attention mechanism.

\item \textbf{FAME}~\cite{liu2020fast} - FAME is a random projection-based network embedding for AMHENs, which uses spectral graph transformation to capture meta-paths, and significantly improves efficiency through random projection. 

\item \textbf{HGSL}~\cite{zhao2021heterogeneous} - HGSL is a state-of-the-art heterogeneous GNN, which jointly performs heterogeneous graph structure learning and GNN parameter learning for classification.  

\item \textbf{DualHGNN}~\cite{xue2021multiplex} - DualHGCN uses dual hypergraph convolutional network to learn node embeddings for multiplex bipartite networks. 

\end{itemize}

The network types handled by the competitor methods are summarized in Table~\ref{tab:method} in the supplemental material.

\vspace{-0mm}
\subsection{Experimental Setting}



For baseline implementations, we either leverage OpenHGNN\footnote{\url{https://github.com/BUPT-GAMMA/OpenHGNN}} or use the released source code for evaluation. In our experiments, we keep $d$=200 for all compared methods. Others hyperparameter settings are considered according to their original papers. For our \system, we set the number of convolution layers $l$ to 2. For fair comparison, we uniformly set the number of training rounds to 500 for link prediction and the number of training rounds to 200 for node classification. 
More detailed experimental settings can be found in the supplementary material.

\vspace{-0mm}
\subsection{Link Prediction}
We first evaluate the model performance by comparing our \system with fifteen baselines on link prediction task in an unsupervised learning manner. 
The results are shown in Table~\ref{table_linkprediction}, where the best is shown in bold. The first seven baselines are homogeneous or heterogeneous network embedding methods, and the last eight are multiplex network embedding methods. 

We can see that \system significantly outperforms all baselines in terms of all evaluation metrics on five datasets. 
Specifically, \system achieves average gains of 5.68\% F1 score in comparison to the best performed GNN baselines across all datasets (\ie, FAME, MAGNN and HPN). 
Our \system realizes a high accuracy of more than 96\% on three datasets (Alibaba, Amazon, and IMDB), especially more than 99\% prediction performance on Alibaba network. 
This is because \system automatically captures effective multi-relational topological structures through multiplex relation aggregation and multilayer graph convolution on the generated meta-paths across multiplex relations.  
Especially, compared with GATNE and MAGNN, our model has achieved better results, showing the ability of our model in automatically capturing meta-paths compared with manually setting meta-paths.
FAME that use spectral graph transformation achieving the second best performance on most datasets also verifies the ability of multiplex relation aggregation to automatically capture useful heterogeneous meta-paths. 
However, \system obtains better performance than FAME on all networks as \system learns meaning node representations for AMHENs using multilayer graph convolution in a learning manner.  
Additionally, \system also shows significant performance advantages on general heterogeneous networks (\eg, IMDB and DBLP). 
This may be because our \system uses a weighted approach to differentiate the effects of different types of relations on node representation, which cannot be achieved by traditional meta-path sampling.


\begin{table*}[t]
\begin{center}
\vspace{-0mm}
\caption{Node classification performance comparison of different methods on four datasets}
\vspace{-2mm}
\label{tab:node_calssfication}
\begin{threeparttable}
\setlength{\tabcolsep}{0.7mm}{}
\begin{tabular}{c|cc|cc|cc|cc}
\toprule
\multirow{2}{*}{Method} & \multicolumn{2}{c|}{AMiner} & \multicolumn{2}{c|}{Alibaba} & \multicolumn{2}{c|}{IMDB} & \multicolumn{2}{c}{DBLP}\\
  & Macro-F1 & Micro-F1 & Macro-F1 & Micro-F1 & Macro-F1 & Micro-F1 & Macro-F1 & Micro-F1 \\
\midrule
node2vec & 0.522 (0.0032) & 0.532 (0.0051) & 0.238 (0.0125) & 0.347 (0.0093) & 0.363 (0.0237) & 0.382 (0.0703) & 0.352 (0.0103) & 0.351 (0.0112)\\
RandNE  & 0.641 (0.0074) & 0.672 (0.0064) & 0.319 (0.0170) & 0.358 (0.0093) & 0.373 (0.0143) & 0.392 (0.0185) & 0.351 (0.0153) & 0.372 (0.0150) \\
FastRP  & 0.650 (0.0086) & 0.690 (0.0074) & 0.301 (0.0180) & 0.392 (0.0119) & 0.363 (0.0236) & 0.381 (0.0140) & 0.343 (0.0201) & 0.375 (0.0199) \\
MNE  & 0.643 (0.0069) & 0.686 (0.0045) & 0.289 (0.0155) & 0.390 (0.0021) & 0.374 (0.0153) & 0.382 (0.0680) & 0.366 (0.0117) & 0.384 (0.0109) \\
GATNE  & OOT & OOT & 0.291 (0.0086) & 0.390 (0.0014) & 0.369 (0.0132) & 0.333 (0.0005) & OOT & OOT \\
DMGI  & 0.473 (0.0155) & 0.626 (0.0093) & 0.220 (0.0214) & 0.392 (0.0026) & 0.548 (0.0190) & 0.544 (0.0189) &0.781 (0.0303) & 0.787 (0.0235)\\ 
FAME  & 0.722 (0.0114) & 0.727 (0.0091) & 0.323 (0.0154) & 0.393 (0.0060) & 0.593 (0.0135) & 0.594 (0.0143) & 0.842 (0.0183) & 0.868 (0.0127)\\
DualHGNN  & / & / & 0.347 (0.0114) & 0.402 (0.0127) & / & / & / & / \\\midrule
SGC  & 0.516 (0.0047) & 0.587 (0.0157) & 0.286 (0.0231) & 0.361 (0.0175) & 0.489 (0.0106) & 0.563 (0.0133) & 0.622 (0.0009) & 0.623 (0.0009)\\ 
AM-GCN  & 0.702 (0.0175) & 0.713 (0.0223) & 0.307 (0.0232) & 0.399 (0.0156) & 0.610 (0.0021) & 0.640 (0.0013) & 0.867 (0.0105) & 0.878 (0.0112)\\ 
R-GCN & 0.690 (0.0078) & 0.692 (0.0106) & 0.265 (0.0326) & 0.381 (0.0125) & 0.544 (0.0172) & 0.572 (0.0145) & 0.862 (0.0053) & 0.870 (0.0070)\\
HAN  & 0.690 (0.0149) & 0.726 (0.0086) & 0.275 (0.0327) & 0.392 (0.0081) & 0.552 (0.0112) & 0.568 (0.0078) & 0.806 (0.0078) & 0.813 (0.0100)\\
NARS & 0.722 (0.0103) & 0.721 (0.0097) & 0.297 (0.0201) & 0.392 (0.0195) & 0.565 (0.0037) & 0.574 (0.0048) & 0.794 (0.0255) & 0.804 (0.0320)\\ 
MAGNN  &  0.755 (0.0105) & 0.757 (0.0133) & 0.348 (0.0488) & 0.398 (0.0405) & 0.614 (0.0073) & 0.615 (0.0089) & 0.881 (0.0284) & 0.895 (0.0396)\\ 
HPN  &  0.710 (0.0612) & 0.732 (0.0490) & 0.263 (0.0346) & 0.392 (0.0405) & 0.578 (0.0023) & 0.584 (0.0021) & 0.822 (0.0201) & 0.830 (0.0201)\\
GTN  & OOM & OOM & 0.255 (0.0420) & 0.392 (0.0071) & 0.615 (0.0108) & 0.616 (0.0093) & 0.852 (0.0137) & 0.868 (0.0125)\\ 
HGSL  &  0.754 (0.0100) & 0.758 (0.0103) & 0.338 (0.0121) & 0.398 (0.0238) & 0.620 (0.0048) & 0.638 (0.0030) & 0.893 (0.0284) & 0.902 (0.0396) \\\hline
\textbf{\system}  & \textbf{0.868 (0.0160)} & \textbf{0.875 (0.0200)} & \textbf{0.351 (0.0204)} & \textbf{0.458 (0.0160)} & \textbf{0.764 (0.0145)} & \textbf{0.782 (0.0138)} & \textbf{0.945 (0.0221)} & \textbf{0.952 (0.0203)} \\ 
\bottomrule
\end{tabular}
\begin{tablenotes}
        \item OOT: Out Of Time (36 hours), OOM: Out Of Memory. The standard deviations are reported in the parentheses. 
\end{tablenotes}
\end{threeparttable}
\end{center}
\vspace{-3mm}
\end{table*}

\begin{figure}[h]
    \begin{center}
    \vspace{-0mm}
    \hspace{-2mm}
    \subfigure[Macro-F1 score]{
    \label{fig:ablation-Ma-F1}
    \includegraphics[width=0.233\textwidth]{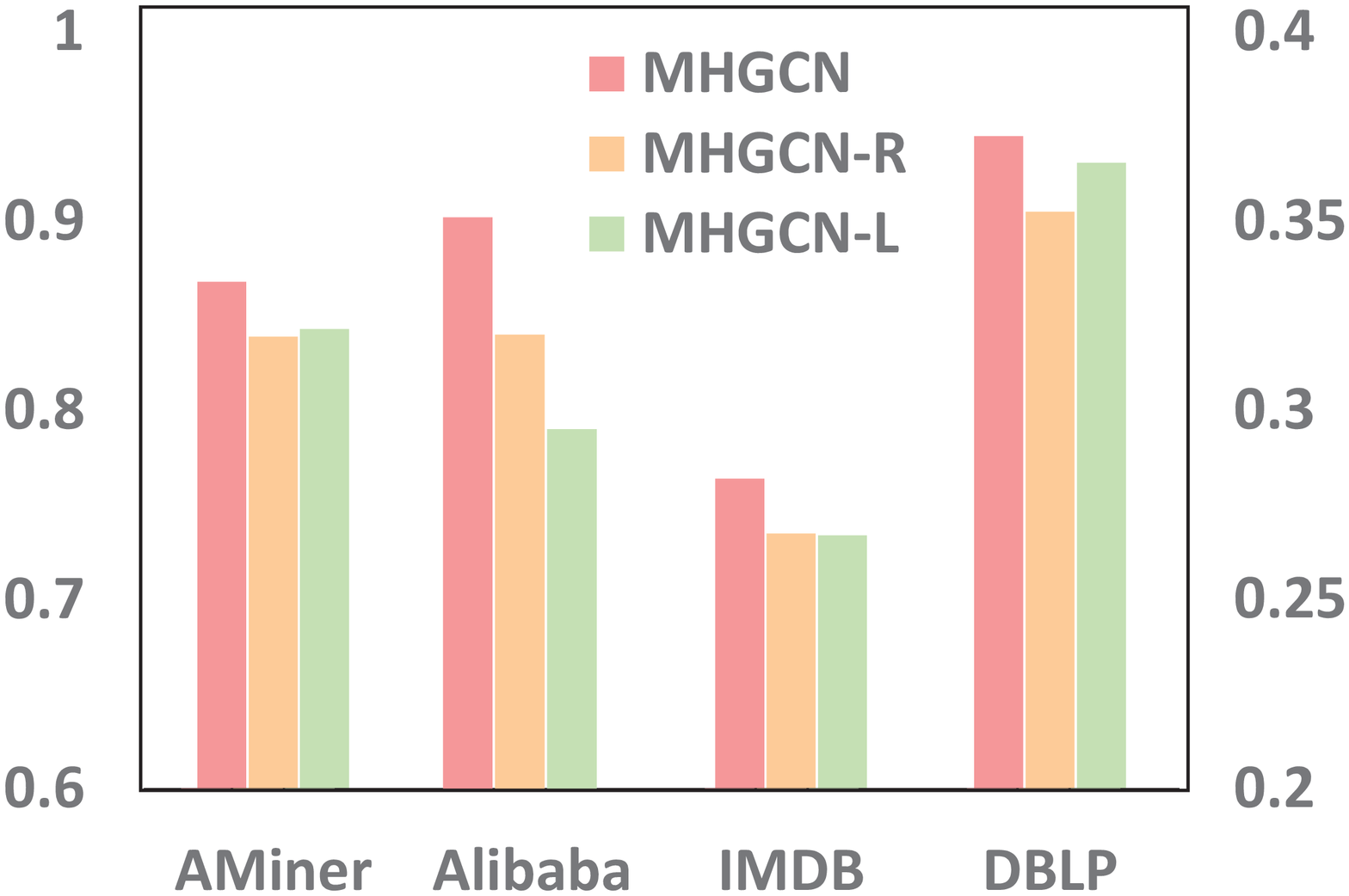}
    }
    \hspace{-2mm}
    \subfigure[Micro-F1 score]{
    \label{fig:ablation-Mi-F1}
    \includegraphics[width=0.233\textwidth]{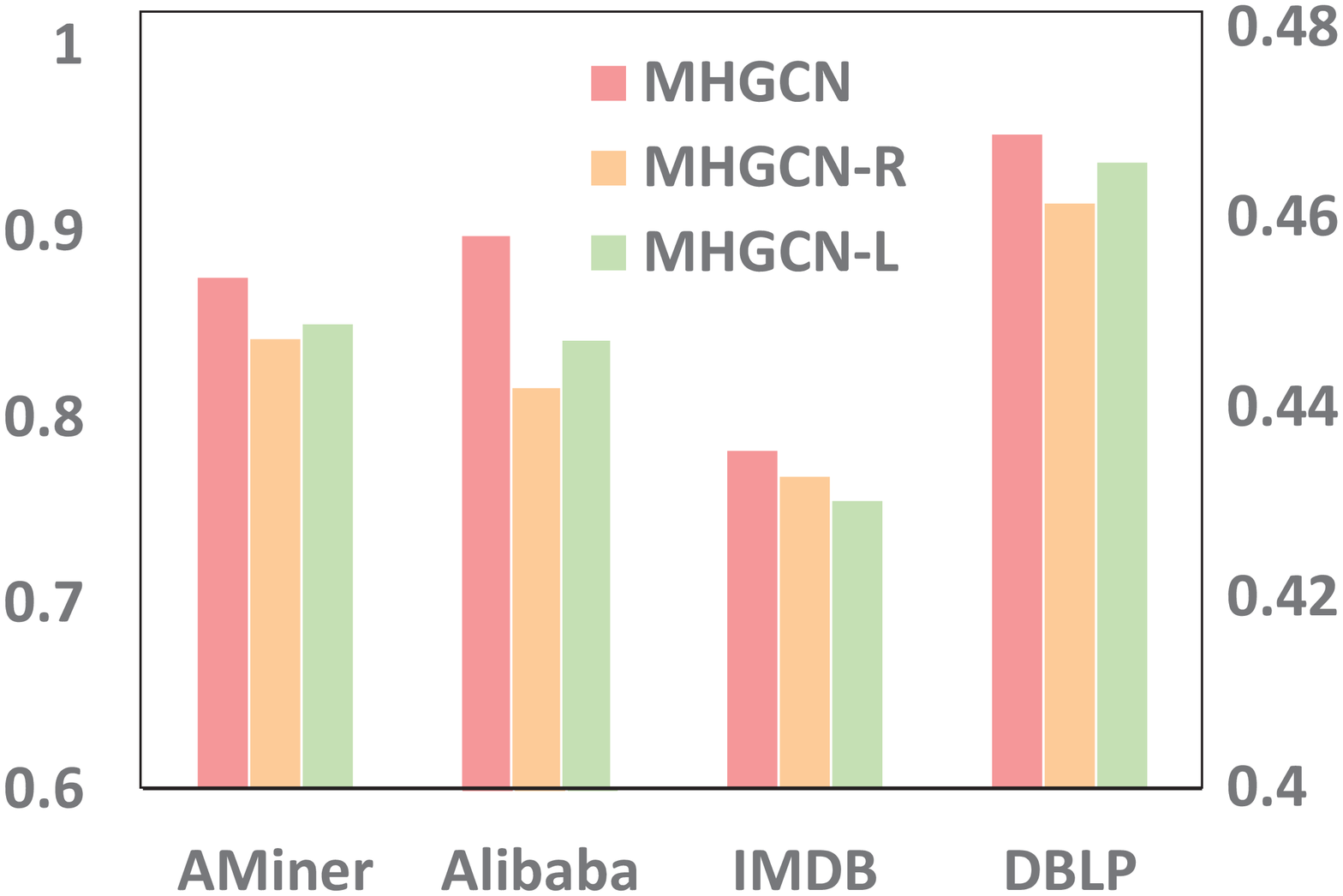}
    }
    \vspace{-4mm}
    \caption{Experimental results of ablation study}
    \label{fig: ablation}
    \vspace{-4mm}
    \end{center}
\end{figure}


\vspace{-0mm}
\subsection{Node Classification}
We next evaluate the effectiveness of our model on the node classification task compared with state-of-the-art methods. 
The results are shown in Table~\ref{tab:node_calssfication}, where the best is shown in bold. 
The first eight baselines are unsupervised embedding methods, and the rest are semi-supervised embedding methods. 

As we see, \system also achieves state-of-the-art performance on all tested networks. Specifically, our \system achieves average 11.22\% and 14.49\% improvement over state-of-the-art GNN model HGSL across all datasets in terms of Macro-F1 and Micro-F1, respectively. 
Considering that the performance gain in node classification task reported in some recent works~\cite{fu2020magnn,zhao2021heterogeneous} is usually around 2-4\%, this performance improvement achieved by our \system is significant. 
Furthermore, we also observe that \system performs much better than competitor methods on general heterogeneous network with multi-typed nodes (\eg, IMDB and AMiner), achieving 23.23\% and 22.19\% improvement in Macro-F1 and Micro-F1 on IMDB network. 
The possible reason is that our \system effectively learns node representations for classification by exploring all meta-path interactions across multiple relations with different importance (\ie, weights), which is ignored by the heterogeneous network embedding approaches based on manually setting meta-path sampling.

\begin{figure*}[h]
    \begin{center}
    \vspace{-0mm}
    \hspace{-3mm}
    \subfigure[Macro-F1 score \wrt. \#layers]{
    \label{fig:layer-Ma-F1}
    \includegraphics[width=0.33\textwidth]{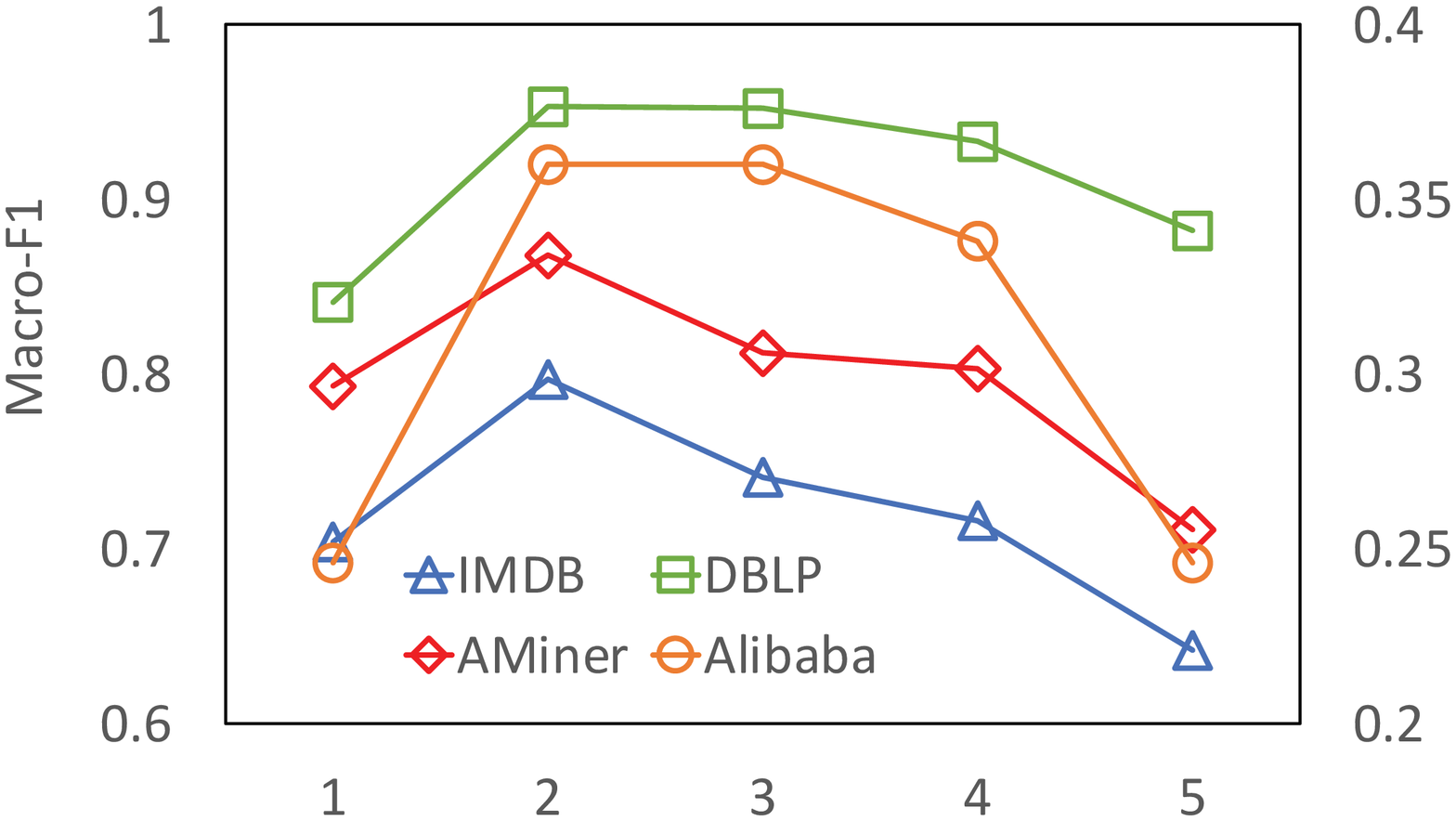}
    }
    \hspace{-2mm}
    \subfigure[Macro-F1 score \wrt. dimension $d$]{
    \label{fig:dimension-Ma-F1}
    \includegraphics[width=0.33\textwidth]{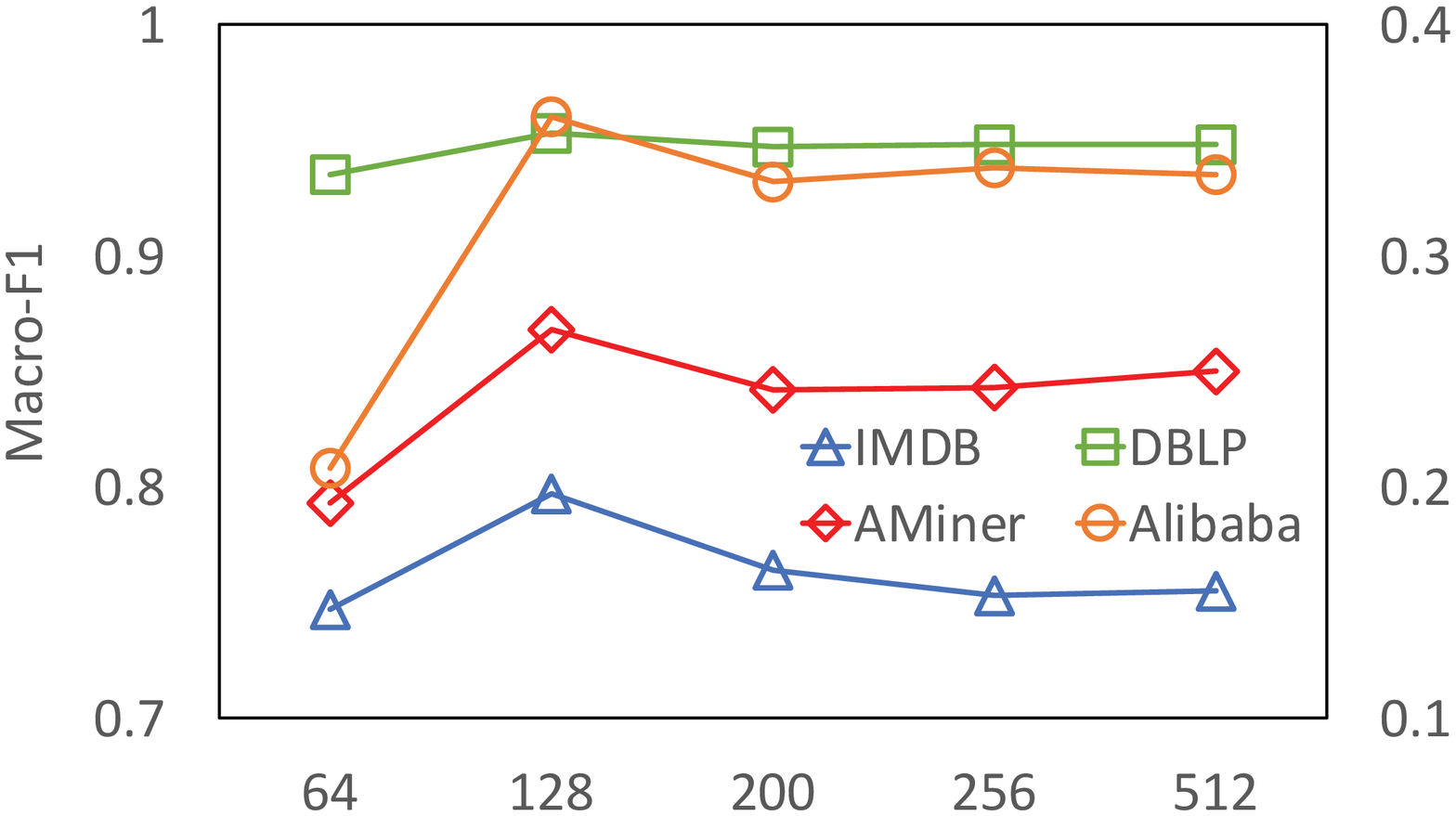}
    }
    \hspace{-2mm}
    \subfigure[Macro-F1 score \wrt. \#rounds]{
    \label{fig:round-Ma-F1}
    \includegraphics[width=0.33\textwidth]{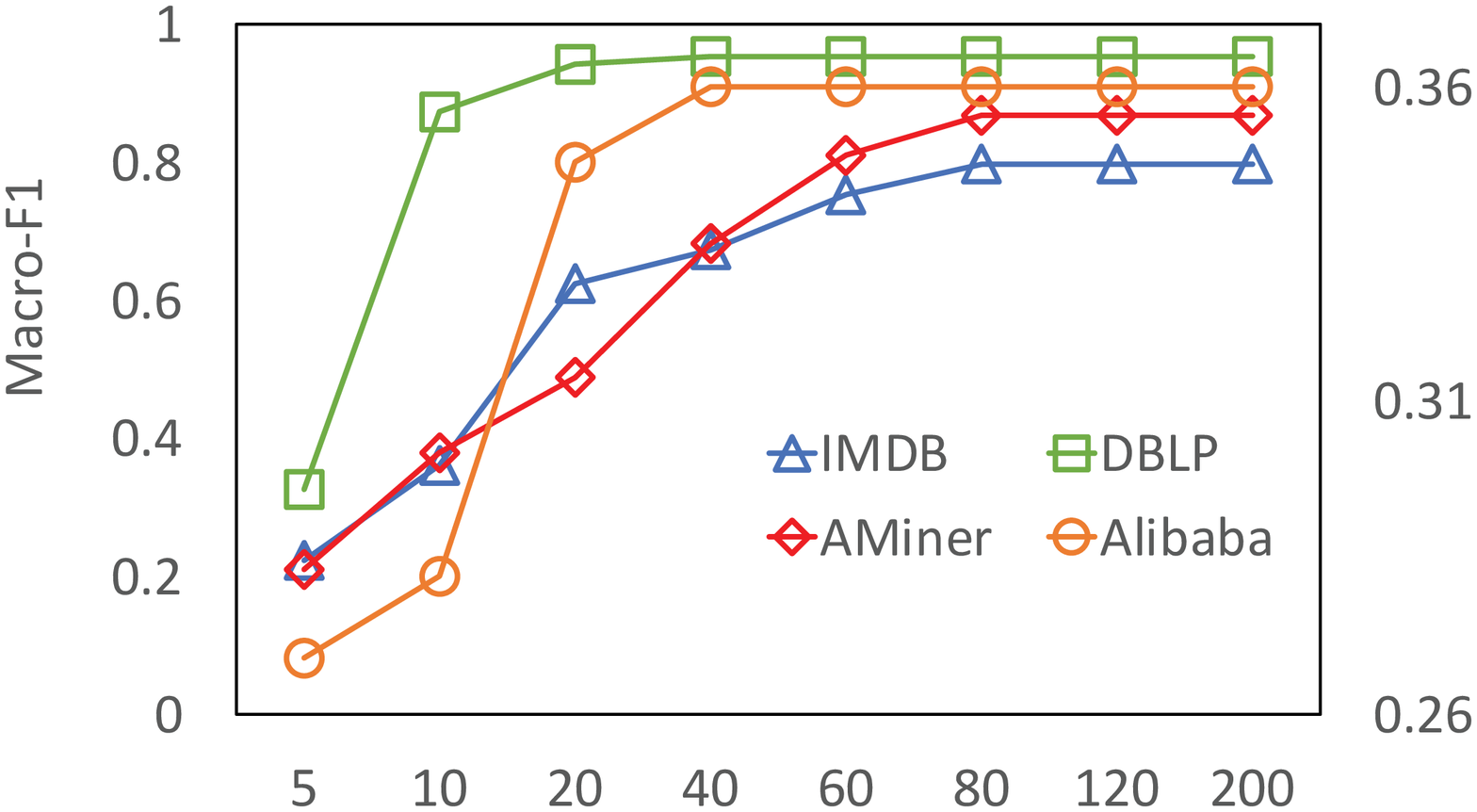}
    }
    \vspace{-4mm}
    \caption{Hyperparameter impact study of the proposed method \wrt. \#layers, dimension $d$, and \#rounds.} 
    \label{fig:sensitive}
    \vspace{-3mm}
    \end{center}
\end{figure*}

\subsection{Ablation Study}

To validate the effectiveness of each component of our model, we further conduct experiments on different \system variations. Here \varr does not consider the importance of different relations, that is, we set the weights $\beta_r$ to 1; \varl uses only a two-layer GCN to obtain the embedding, so it can only capture the 2-length meta-paths. We report the results of ablation study on four datasets for node classification in Figure~\ref{fig: ablation}, where the performance on Alibaba refers to the right-ordinate axis. 

It can be seen from the results that the two key components both contribute to performance improvement of our \system. 
The comparison between \varr and \system highlights the effectiveness of the importance of different relations. 
We can observe that \varr performs worse than \system on all datasets in terms of both Macro-F1 and Micro-F1 metrics, reducing 9.68\% performance in Macro-F1 score on Alibaba, which demonstrates the crucial role of our designed multiplex relation aggregation module in capturing the importance of different relations for node representation learning. 
The comparison between \varl and \system reflects the importance of our multilayer graph convolution module. 
Compared with \varl, \system improves 2.97\%, 18.98\%, 4.09\% and 1.51\% over \varl in terms of Macro-F1 on AMiner, Alibaba, IMDB, and DBLP, respectively.
This indicates that our proposed multilayer graph convolution module effectively captures useful meta-paths of different lengths across multiplex relations.

\vspace{-0mm}
\subsection{Parameter Sensitivity}

We conduct hyperparameter study in our new \system in terms of key parameters, \ie the number of graph layers $l$, the hidden dimension size of embeddings $d$, and the number of training rounds. We report Macro-F1 score on node classification task with different parameter settings on four datasets in Figure~\ref{fig:sensitive}. Notice that the performance on Alibaba refers to the ordinate on the right.


From the results shown in Figure~\ref{fig:layer-Ma-F1}, we can observe that the best performance can be achieved with two layers. This observation indicates that considering the meta-path interactions with two-hops is sufficient to capture the node dependencies in the graph. Performing the message passing across more graph layers may involve noisy information for node representation. With the growth of GCN layers, the representation of nodes would be flattened after multiple convolutions, resulting in performance degradation. Additionally, we can notice that the increasing of embedding dimension first brings benefits for the performance improvement, and then leads to the performance degradation. The best prediction accuracy can be achieved with the setting of $d=128$. This is because the features of all nodes are compressed into a small embedding space when dimension $d$ is small, thus it is difficult to retain the characteristics proximities of all node pairs. 
Conversely, a larger dimension would also flatten the distance between all node embeddings. Figure~\ref{fig:round-Ma-F1} illustrates the performance of our \system with respect to the number of training rounds in learning model weights. 
We can find that our \system can converge quickly and efficiently achieve stable performance within 80 rounds on all tested datasets.




\section{Conclusion}

In this paper, we propose an embedding model \system for attributed multiplex heterogeneous networks. Our model mainly includes two key components: multiplex relation aggregation and multilayer graph convolution module. Through multiplex relation aggregation, \system can distinguish the importance of the relations between different nodes in multiplex heterogeneous networks. Through multilayer graph convolution module, \system can automatically capture the short and long meta-path interactions across multi-relations, and learn meaning node embeddings with model parameter learning during training phase. Experiments results on five real-world heterogeneous networks show the superiority of the proposed \system in different graph representation tasks.


\begin{acks}
This work is partially supported by the National Natural Science Foundation of China under grant Nos. 62176243, 62072288, 61773331 and 41927805, and the National Key Research and Development Program of China under grant Nos. 2018AAA0100602 and 2019YFC1509100.
\end{acks}

\balance 
\bibliographystyle{ACM-Reference-Format}
\bibliography{kdd2022}


\begin{thebibliography}{43}


\ifx \showCODEN    \undefined \def \showCODEN     #1{\unskip}     \fi
\ifx \showDOI      \undefined \def \showDOI       #1{#1}\fi
\ifx \showISBNx    \undefined \def \showISBNx     #1{\unskip}     \fi
\ifx \showISBNxiii \undefined \def \showISBNxiii  #1{\unskip}     \fi
\ifx \showISSN     \undefined \def \showISSN      #1{\unskip}     \fi
\ifx \showLCCN     \undefined \def \showLCCN      #1{\unskip}     \fi
\ifx \shownote     \undefined \def \shownote      #1{#1}          \fi
\ifx \showarticletitle \undefined \def \showarticletitle #1{#1}   \fi
\ifx \showURL      \undefined \def \showURL       {\relax}        \fi
\providecommand\bibfield[2]{#2}
\providecommand\bibinfo[2]{#2}
\providecommand\natexlab[1]{#1}
\providecommand\showeprint[2][]{arXiv:#2}

\bibitem[\protect\citeauthoryear{Cen, Zou, Zhang, Yang, Zhou, and Tang}{Cen
  et~al\mbox{.}}{2019}]%
        {kdd19GATNE}
\bibfield{author}{\bibinfo{person}{Yukuo Cen}, \bibinfo{person}{Xu Zou},
  \bibinfo{person}{Jianwei Zhang}, \bibinfo{person}{Hongxia Yang},
  \bibinfo{person}{Jingren Zhou}, {and} \bibinfo{person}{Jie Tang}.}
  \bibinfo{year}{2019}\natexlab{}.
\newblock \showarticletitle{Representation Learning for Attributed Multiplex
  Heterogeneous Network}. In \bibinfo{booktitle}{\emph{KDD}}.
  \bibinfo{pages}{1358--1368}.
\newblock


\bibitem[\protect\citeauthoryear{Chen, Sultan, Tian, Chen, and Skiena}{Chen
  et~al\mbox{.}}{2019}]%
        {DBLP:conf/cikm19_fastRP}
\bibfield{author}{\bibinfo{person}{Haochen Chen}, \bibinfo{person}{Syed~Fahad
  Sultan}, \bibinfo{person}{Yingtao Tian}, \bibinfo{person}{Muhao Chen}, {and}
  \bibinfo{person}{Steven Skiena}.} \bibinfo{year}{2019}\natexlab{}.
\newblock \showarticletitle{Fast and Accurate Network Embeddings via Very
  Sparse Random Projection}. In \bibinfo{booktitle}{\emph{CIKM}}.
  \bibinfo{pages}{399--408}.
\newblock


\bibitem[\protect\citeauthoryear{Chen, Yin, Wang, Wang, Nguyen, and Li}{Chen
  et~al\mbox{.}}{2018}]%
        {chen2018pme}
\bibfield{author}{\bibinfo{person}{Hongxu Chen}, \bibinfo{person}{Hongzhi Yin},
  \bibinfo{person}{Weiqing Wang}, \bibinfo{person}{Hao Wang},
  \bibinfo{person}{Quoc Viet~Hung Nguyen}, {and} \bibinfo{person}{Xue Li}.}
  \bibinfo{year}{2018}\natexlab{}.
\newblock \showarticletitle{PME: projected metric embedding on heterogeneous
  networks for link prediction}. In \bibinfo{booktitle}{\emph{KDD}}.
  \bibinfo{pages}{1177--1186}.
\newblock


\bibitem[\protect\citeauthoryear{Dong, Chawla, and Swami}{Dong
  et~al\mbox{.}}{2017}]%
        {dong2017metapath2vec}
\bibfield{author}{\bibinfo{person}{Yuxiao Dong}, \bibinfo{person}{Nitesh~V
  Chawla}, {and} \bibinfo{person}{Ananthram Swami}.}
  \bibinfo{year}{2017}\natexlab{}.
\newblock \showarticletitle{metapath2vec: Scalable representation learning for
  heterogeneous networks}. In \bibinfo{booktitle}{\emph{KDD}}.
  \bibinfo{pages}{135--144}.
\newblock


\bibitem[\protect\citeauthoryear{Fu, Zhang, Meng, and King}{Fu
  et~al\mbox{.}}{2020}]%
        {fu2020magnn}
\bibfield{author}{\bibinfo{person}{Xinyu Fu}, \bibinfo{person}{Jiani Zhang},
  \bibinfo{person}{Ziqiao Meng}, {and} \bibinfo{person}{Irwin King}.}
  \bibinfo{year}{2020}\natexlab{}.
\newblock \showarticletitle{Magnn: Metapath aggregated graph neural network for
  heterogeneous graph embedding}. In \bibinfo{booktitle}{\emph{WWW}}.
  \bibinfo{pages}{2331--2341}.
\newblock


\bibitem[\protect\citeauthoryear{Grover and Leskovec}{Grover and
  Leskovec}{2016}]%
        {grover2016node2vec}
\bibfield{author}{\bibinfo{person}{Aditya Grover} {and} \bibinfo{person}{Jure
  Leskovec}.} \bibinfo{year}{2016}\natexlab{}.
\newblock \showarticletitle{node2vec: Scalable feature learning for networks}.
  In \bibinfo{booktitle}{\emph{KDD}}. ACM, \bibinfo{pages}{855--864}.
\newblock


\bibitem[\protect\citeauthoryear{Hamilton, Ying, and Leskovec}{Hamilton
  et~al\mbox{.}}{2017}]%
        {William2017GraphSAGE}
\bibfield{author}{\bibinfo{person}{William~L. Hamilton}, \bibinfo{person}{Rex
  Ying}, {and} \bibinfo{person}{Jure Leskovec}.}
  \bibinfo{year}{2017}\natexlab{}.
\newblock \showarticletitle{Inductive Representation Learning on Large Graphs}.
  In \bibinfo{booktitle}{\emph{NeurIPS}}. \bibinfo{pages}{1025–1035}.
\newblock


\bibitem[\protect\citeauthoryear{He, Chen, Wang, Jameel, Yu, and Xu}{He
  et~al\mbox{.}}{2021}]%
        {he2021click}
\bibfield{author}{\bibinfo{person}{Li He}, \bibinfo{person}{Hongxu Chen},
  \bibinfo{person}{Dingxian Wang}, \bibinfo{person}{Shoaib Jameel},
  \bibinfo{person}{Philip Yu}, {and} \bibinfo{person}{Guandong Xu}.}
  \bibinfo{year}{2021}\natexlab{}.
\newblock \showarticletitle{Click-Through Rate Prediction with Multi-Modal
  Hypergraphs}. In \bibinfo{booktitle}{\emph{CIKM}}. \bibinfo{pages}{690--699}.
\newblock


\bibitem[\protect\citeauthoryear{He, Deng, Wang, Li, Zhang, and Wang}{He
  et~al\mbox{.}}{2020}]%
        {he2020lightgcn}
\bibfield{author}{\bibinfo{person}{Xiangnan He}, \bibinfo{person}{Kuan Deng},
  \bibinfo{person}{Xiang Wang}, \bibinfo{person}{Yan Li},
  \bibinfo{person}{Yongdong Zhang}, {and} \bibinfo{person}{Meng Wang}.}
  \bibinfo{year}{2020}\natexlab{}.
\newblock \showarticletitle{Lightgcn: Simplifying and powering graph
  convolution network for recommendation}. In
  \bibinfo{booktitle}{\emph{SIGIR}}. \bibinfo{pages}{639--648}.
\newblock


\bibitem[\protect\citeauthoryear{Hu, Fang, and Shi}{Hu et~al\mbox{.}}{2019}]%
        {kdd19HeGAN}
\bibfield{author}{\bibinfo{person}{Binbin Hu}, \bibinfo{person}{Yuan Fang},
  {and} \bibinfo{person}{Chuan Shi}.} \bibinfo{year}{2019}\natexlab{}.
\newblock \showarticletitle{Adversarial Learning on Heterogeneous Information
  Networks}. In \bibinfo{booktitle}{\emph{KDD}}. \bibinfo{pages}{120--129}.
\newblock


\bibitem[\protect\citeauthoryear{Hu, Dong, Wang, and Sun}{Hu
  et~al\mbox{.}}{2020}]%
        {hu2020heterogeneous}
\bibfield{author}{\bibinfo{person}{Ziniu Hu}, \bibinfo{person}{Yuxiao Dong},
  \bibinfo{person}{Kuansan Wang}, {and} \bibinfo{person}{Yizhou Sun}.}
  \bibinfo{year}{2020}\natexlab{}.
\newblock \showarticletitle{Heterogeneous graph transformer}. In
  \bibinfo{booktitle}{\emph{WWW}}. \bibinfo{pages}{2704--2710}.
\newblock


\bibitem[\protect\citeauthoryear{Huang}{Huang}{2021}]%
        {2021recent}
\bibfield{author}{\bibinfo{person}{Chao Huang}.}
  \bibinfo{year}{2021}\natexlab{}.
\newblock \showarticletitle{Recent advances in heterogeneous relation learning
  for recommendation}.
\newblock \bibinfo{journal}{\emph{arXiv preprint arXiv:2110.03455}}
  (\bibinfo{year}{2021}).
\newblock


\bibitem[\protect\citeauthoryear{Huang, Xu, Xu, Dai, Xiao, Lu, Bo, Xing, Lai,
  and Ye}{Huang et~al\mbox{.}}{2021}]%
        {2021knowledge}
\bibfield{author}{\bibinfo{person}{Chao Huang}, \bibinfo{person}{Huance Xu},
  \bibinfo{person}{Yong Xu}, \bibinfo{person}{Peng Dai},
  \bibinfo{person}{Lianghao Xiao}, \bibinfo{person}{Mengyin Lu},
  \bibinfo{person}{Liefeng Bo}, \bibinfo{person}{Hao Xing},
  \bibinfo{person}{Xiaoping Lai}, {and} \bibinfo{person}{Yanfang Ye}.}
  \bibinfo{year}{2021}\natexlab{}.
\newblock \showarticletitle{Knowledge-aware coupled graph neural network for
  social recommendation}. In \bibinfo{booktitle}{\emph{AAAI}}.
\newblock


\bibitem[\protect\citeauthoryear{Ji, Wang, Shi, Wang, and Yu}{Ji
  et~al\mbox{.}}{2021}]%
        {ji2021heterogeneous}
\bibfield{author}{\bibinfo{person}{Houye Ji}, \bibinfo{person}{Xiao Wang},
  \bibinfo{person}{Chuan Shi}, \bibinfo{person}{Bai Wang}, {and}
  \bibinfo{person}{Philip Yu}.} \bibinfo{year}{2021}\natexlab{}.
\newblock \showarticletitle{Heterogeneous Graph Propagation Network}.
\newblock \bibinfo{journal}{\emph{TKDE}} (\bibinfo{year}{2021}).
\newblock


\bibitem[\protect\citeauthoryear{Jing, Park, and Tong}{Jing
  et~al\mbox{.}}{2021}]%
        {jing2021hdmi}
\bibfield{author}{\bibinfo{person}{Baoyu Jing}, \bibinfo{person}{Chanyoung
  Park}, {and} \bibinfo{person}{Hanghang Tong}.}
  \bibinfo{year}{2021}\natexlab{}.
\newblock \showarticletitle{Hdmi: High-order deep multiplex infomax}. In
  \bibinfo{booktitle}{\emph{The Web Conference}}. \bibinfo{pages}{2414--2424}.
\newblock


\bibitem[\protect\citeauthoryear{Kipf and Welling}{Kipf and Welling}{2017}]%
        {iclr17KipfGCN}
\bibfield{author}{\bibinfo{person}{Thomas~N. Kipf} {and} \bibinfo{person}{Max
  Welling}.} \bibinfo{year}{2017}\natexlab{}.
\newblock \showarticletitle{Semi-Supervised Classification with Graph
  Convolutional Networks}. In \bibinfo{booktitle}{\emph{ICLR}}.
\newblock


\bibitem[\protect\citeauthoryear{Li, Shen, Jiao, Pan, Zou, Meng, Yao, and
  Bu}{Li et~al\mbox{.}}{2020}]%
        {li2020hierarchical}
\bibfield{author}{\bibinfo{person}{Zhao Li}, \bibinfo{person}{Xin Shen},
  \bibinfo{person}{Yuhang Jiao}, \bibinfo{person}{Xuming Pan},
  \bibinfo{person}{Pengcheng Zou}, \bibinfo{person}{Xianling Meng},
  \bibinfo{person}{Chengwei Yao}, {and} \bibinfo{person}{Jiajun Bu}.}
  \bibinfo{year}{2020}\natexlab{}.
\newblock \showarticletitle{Hierarchical bipartite graph neural networks:
  Towards large-scale e-commerce applications}. In
  \bibinfo{booktitle}{\emph{ICDE}}. IEEE, \bibinfo{pages}{1677--1688}.
\newblock


\bibitem[\protect\citeauthoryear{Liu, Chen, Yeung, Suzumura, and Chen}{Liu
  et~al\mbox{.}}{2017}]%
        {liu2017PMNE}
\bibfield{author}{\bibinfo{person}{Weiyi Liu}, \bibinfo{person}{Pin-Yu Chen},
  \bibinfo{person}{Sailung Yeung}, \bibinfo{person}{Toyotaro Suzumura}, {and}
  \bibinfo{person}{Lingli Chen}.} \bibinfo{year}{2017}\natexlab{}.
\newblock \showarticletitle{Principled multilayer network embedding}. In
  \bibinfo{booktitle}{\emph{ICDMW}}. IEEE, \bibinfo{pages}{134--141}.
\newblock


\bibitem[\protect\citeauthoryear{Liu, Huang, Yu, and Dong}{Liu
  et~al\mbox{.}}{2021}]%
        {liu2021motif}
\bibfield{author}{\bibinfo{person}{Zhijun Liu}, \bibinfo{person}{Chao Huang},
  \bibinfo{person}{Yanwei Yu}, {and} \bibinfo{person}{Junyu Dong}.}
  \bibinfo{year}{2021}\natexlab{}.
\newblock \showarticletitle{Motif-preserving dynamic attributed network
  embedding}. In \bibinfo{booktitle}{\emph{WWW}}. \bibinfo{pages}{1629--1638}.
\newblock


\bibitem[\protect\citeauthoryear{Liu, Huang, Yu, Fan, and Dong}{Liu
  et~al\mbox{.}}{2020}]%
        {liu2020fast}
\bibfield{author}{\bibinfo{person}{Zhijun Liu}, \bibinfo{person}{Chao Huang},
  \bibinfo{person}{Yanwei Yu}, \bibinfo{person}{Baode Fan}, {and}
  \bibinfo{person}{Junyu Dong}.} \bibinfo{year}{2020}\natexlab{}.
\newblock \showarticletitle{Fast Attributed Multiplex Heterogeneous Network
  Embedding}. In \bibinfo{booktitle}{\emph{CIKM}}. \bibinfo{pages}{995--1004}.
\newblock


\bibitem[\protect\citeauthoryear{Long, Huang, Xu, Xu, Dai, Xia, and Bo}{Long
  et~al\mbox{.}}{2021}]%
        {long2021social}
\bibfield{author}{\bibinfo{person}{Xiaoling Long}, \bibinfo{person}{Chao
  Huang}, \bibinfo{person}{Yong Xu}, \bibinfo{person}{Huance Xu},
  \bibinfo{person}{Peng Dai}, \bibinfo{person}{Lianghao Xia}, {and}
  \bibinfo{person}{Liefeng Bo}.} \bibinfo{year}{2021}\natexlab{}.
\newblock \showarticletitle{Social Recommendation with Self-Supervised
  Metagraph Informax Network}. In \bibinfo{booktitle}{\emph{CIKM}}.
  \bibinfo{pages}{1160--1169}.
\newblock


\bibitem[\protect\citeauthoryear{Lu, Shi, Hu, and Liu}{Lu
  et~al\mbox{.}}{2019}]%
        {lu2019relation}
\bibfield{author}{\bibinfo{person}{Yuanfu Lu}, \bibinfo{person}{Chuan Shi},
  \bibinfo{person}{Linmei Hu}, {and} \bibinfo{person}{Zhiyuan Liu}.}
  \bibinfo{year}{2019}\natexlab{}.
\newblock \showarticletitle{Relation structure-aware heterogeneous information
  network embedding}. In \bibinfo{booktitle}{\emph{AAAI}}.
  \bibinfo{pages}{4456--4463}.
\newblock


\bibitem[\protect\citeauthoryear{Park, Kim, Han, and Yu}{Park
  et~al\mbox{.}}{2020}]%
        {park2019DMGI}
\bibfield{author}{\bibinfo{person}{Chanyoung Park}, \bibinfo{person}{Donghyun
  Kim}, \bibinfo{person}{Jiawei Han}, {and} \bibinfo{person}{Hwanjo Yu}.}
  \bibinfo{year}{2020}\natexlab{}.
\newblock \showarticletitle{Unsupervised Attributed Multiplex Network
  Embedding}. In \bibinfo{booktitle}{\emph{AAAI}}. \bibinfo{pages}{5371--5378}.
\newblock


\bibitem[\protect\citeauthoryear{Perozzi, Al{-}Rfou, and Skiena}{Perozzi
  et~al\mbox{.}}{2014}]%
        {DBLP:conf/kdd14Deepwalk}
\bibfield{author}{\bibinfo{person}{Bryan Perozzi}, \bibinfo{person}{Rami
  Al{-}Rfou}, {and} \bibinfo{person}{Steven Skiena}.}
  \bibinfo{year}{2014}\natexlab{}.
\newblock \showarticletitle{DeepWalk: online learning of social
  representations}. In \bibinfo{booktitle}{\emph{KDD}}.
  \bibinfo{pages}{701--710}.
\newblock


\bibitem[\protect\citeauthoryear{Qiu, Dong, Ma, Li, Wang, Wang, and Tang}{Qiu
  et~al\mbox{.}}{2019}]%
        {qiu2019netsmf}
\bibfield{author}{\bibinfo{person}{Jiezhong Qiu}, \bibinfo{person}{Yuxiao
  Dong}, \bibinfo{person}{Hao Ma}, \bibinfo{person}{Jian Li},
  \bibinfo{person}{Chi Wang}, \bibinfo{person}{Kuansan Wang}, {and}
  \bibinfo{person}{Jie Tang}.} \bibinfo{year}{2019}\natexlab{}.
\newblock \showarticletitle{Netsmf: Large-scale network embedding as sparse
  matrix factorization}. In \bibinfo{booktitle}{\emph{WWW}}.
  \bibinfo{pages}{1509--1520}.
\newblock


\bibitem[\protect\citeauthoryear{Schlichtkrull, Kipf, Bloem, Van Den~Berg,
  Titov, and Welling}{Schlichtkrull et~al\mbox{.}}{2018}]%
        {schlichtkrull2018modeling}
\bibfield{author}{\bibinfo{person}{Michael Schlichtkrull},
  \bibinfo{person}{Thomas~N Kipf}, \bibinfo{person}{Peter Bloem},
  \bibinfo{person}{Rianne Van Den~Berg}, \bibinfo{person}{Ivan Titov}, {and}
  \bibinfo{person}{Max Welling}.} \bibinfo{year}{2018}\natexlab{}.
\newblock \showarticletitle{Modeling relational data with graph convolutional
  networks}. In \bibinfo{booktitle}{\emph{European semantic web conference}}.
  Springer, \bibinfo{pages}{593--607}.
\newblock


\bibitem[\protect\citeauthoryear{Shi, Hu, Zhao, and Philip}{Shi
  et~al\mbox{.}}{2018}]%
        {shi2018heterogeneous}
\bibfield{author}{\bibinfo{person}{Chuan Shi}, \bibinfo{person}{Binbin Hu},
  \bibinfo{person}{Wayne~Xin Zhao}, {and} \bibinfo{person}{S~Yu Philip}.}
  \bibinfo{year}{2018}\natexlab{}.
\newblock \showarticletitle{Heterogeneous information network embedding for
  recommendation}.
\newblock \bibinfo{journal}{\emph{TKDE}} \bibinfo{volume}{31},
  \bibinfo{number}{2} (\bibinfo{year}{2018}), \bibinfo{pages}{357--370}.
\newblock


\bibitem[\protect\citeauthoryear{Tang, Qu, Wang, Zhang, Yan, and Mei}{Tang
  et~al\mbox{.}}{2015}]%
        {tang2015line}
\bibfield{author}{\bibinfo{person}{Jian Tang}, \bibinfo{person}{Meng Qu},
  \bibinfo{person}{Mingzhe Wang}, \bibinfo{person}{Ming Zhang},
  \bibinfo{person}{Jun Yan}, {and} \bibinfo{person}{Qiaozhu Mei}.}
  \bibinfo{year}{2015}\natexlab{}.
\newblock \showarticletitle{Line: Large-scale information network embedding}.
  In \bibinfo{booktitle}{\emph{WWW}}. \bibinfo{pages}{1067--1077}.
\newblock


\bibitem[\protect\citeauthoryear{Velickovic, Cucurull, Casanova, Romero,
  Li{\`{o}}, and Bengio}{Velickovic et~al\mbox{.}}{2018}]%
        {DBLP:conf/iclr18GAT}
\bibfield{author}{\bibinfo{person}{Petar Velickovic}, \bibinfo{person}{Guillem
  Cucurull}, \bibinfo{person}{Arantxa Casanova}, \bibinfo{person}{Adriana
  Romero}, \bibinfo{person}{Pietro Li{\`{o}}}, {and} \bibinfo{person}{Yoshua
  Bengio}.} \bibinfo{year}{2018}\natexlab{}.
\newblock \showarticletitle{Graph Attention Networks}. In
  \bibinfo{booktitle}{\emph{ICLR}}.
\newblock


\bibitem[\protect\citeauthoryear{Wang, Ji, Shi, Wang, Ye, Cui, and Yu}{Wang
  et~al\mbox{.}}{2019}]%
        {www2019HAN}
\bibfield{author}{\bibinfo{person}{Xiao Wang}, \bibinfo{person}{Houye Ji},
  \bibinfo{person}{Chuan Shi}, \bibinfo{person}{Bai Wang},
  \bibinfo{person}{Yanfang Ye}, \bibinfo{person}{Peng Cui}, {and}
  \bibinfo{person}{Philip~S Yu}.} \bibinfo{year}{2019}\natexlab{}.
\newblock \showarticletitle{Heterogeneous Graph Attention Network}. In
  \bibinfo{booktitle}{\emph{WWW}}. ACM, \bibinfo{pages}{2022--2032}.
\newblock


\bibitem[\protect\citeauthoryear{Wang, Lu, Shi, Wang, Cui, and Mou}{Wang
  et~al\mbox{.}}{2020a}]%
        {wang2020dynamic}
\bibfield{author}{\bibinfo{person}{Xiao Wang}, \bibinfo{person}{Yuanfu Lu},
  \bibinfo{person}{Chuan Shi}, \bibinfo{person}{Ruijia Wang},
  \bibinfo{person}{Peng Cui}, {and} \bibinfo{person}{Shuai Mou}.}
  \bibinfo{year}{2020}\natexlab{a}.
\newblock \showarticletitle{Dynamic heterogeneous information network embedding
  with meta-path based proximity}.
\newblock \bibinfo{journal}{\emph{TKDE}} (\bibinfo{year}{2020}).
\newblock


\bibitem[\protect\citeauthoryear{Wang, Zhu, Bo, Cui, Shi, and Pei}{Wang
  et~al\mbox{.}}{2020b}]%
        {wang2020gcn}
\bibfield{author}{\bibinfo{person}{Xiao Wang}, \bibinfo{person}{Meiqi Zhu},
  \bibinfo{person}{Deyu Bo}, \bibinfo{person}{Peng Cui}, \bibinfo{person}{Chuan
  Shi}, {and} \bibinfo{person}{Jian Pei}.} \bibinfo{year}{2020}\natexlab{b}.
\newblock \showarticletitle{Am-gcn: Adaptive multi-channel graph convolutional
  networks}. In \bibinfo{booktitle}{\emph{KDD}}. \bibinfo{pages}{1243--1253}.
\newblock


\bibitem[\protect\citeauthoryear{Wei, Huang, Xia, Xu, Zhao, and Yin}{Wei
  et~al\mbox{.}}{2022}]%
        {wei2022contrastive}
\bibfield{author}{\bibinfo{person}{Wei Wei}, \bibinfo{person}{Chao Huang},
  \bibinfo{person}{Lianghao Xia}, \bibinfo{person}{Yong Xu},
  \bibinfo{person}{Jiashu Zhao}, {and} \bibinfo{person}{Dawei Yin}.}
  \bibinfo{year}{2022}\natexlab{}.
\newblock \showarticletitle{Contrastive meta learning with behavior
  multiplicity for recommendation}. In \bibinfo{booktitle}{\emph{KDD}}.
  \bibinfo{pages}{1120--1128}.
\newblock


\bibitem[\protect\citeauthoryear{Wu, Souza, Zhang, Fifty, Yu, et~al\mbox{.}}{Wu
  et~al\mbox{.}}{2019}]%
        {wu2019simplifying}
\bibfield{author}{\bibinfo{person}{Felix Wu}, \bibinfo{person}{Amauri Souza},
  \bibinfo{person}{Tianyi Zhang}, \bibinfo{person}{Christopher Fifty},
  \bibinfo{person}{Tao Yu}, {et~al\mbox{.}}} \bibinfo{year}{2019}\natexlab{}.
\newblock \showarticletitle{Simplifying Graph Convolutional Networks}. In
  \bibinfo{booktitle}{\emph{ICML}}. \bibinfo{pages}{6861--6871}.
\newblock


\bibitem[\protect\citeauthoryear{Wu, Lian, Xu, Wu, and Chen}{Wu
  et~al\mbox{.}}{2020}]%
        {wu2020graph}
\bibfield{author}{\bibinfo{person}{Yongji Wu}, \bibinfo{person}{Defu Lian},
  \bibinfo{person}{Yiheng Xu}, \bibinfo{person}{Le Wu}, {and}
  \bibinfo{person}{Enhong Chen}.} \bibinfo{year}{2020}\natexlab{}.
\newblock \showarticletitle{Graph convolutional networks with markov random
  field reasoning for social spammer detection}. In
  \bibinfo{booktitle}{\emph{AAAI}}, Vol.~\bibinfo{volume}{34}.
  \bibinfo{pages}{1054--1061}.
\newblock


\bibitem[\protect\citeauthoryear{Xia, Huang, Xu, Dai, Zhang, and Bo}{Xia
  et~al\mbox{.}}{2020}]%
        {2020multiplex}
\bibfield{author}{\bibinfo{person}{Lianghao Xia}, \bibinfo{person}{Chao Huang},
  \bibinfo{person}{Yong Xu}, \bibinfo{person}{Peng Dai}, \bibinfo{person}{Bo
  Zhang}, {and} \bibinfo{person}{Liefeng Bo}.} \bibinfo{year}{2020}\natexlab{}.
\newblock \showarticletitle{Multiplex Behavioral Relation Learning for
  Recommendation via Memory Augmented Transformer Network}. In
  \bibinfo{booktitle}{\emph{SIGIR}}. \bibinfo{pages}{2397--2406}.
\newblock


\bibitem[\protect\citeauthoryear{Xue, Yang, Rajan, Jiang, Wei, and Lin}{Xue
  et~al\mbox{.}}{2021}]%
        {xue2021multiplex}
\bibfield{author}{\bibinfo{person}{Hansheng Xue}, \bibinfo{person}{Luwei Yang},
  \bibinfo{person}{Vaibhav Rajan}, \bibinfo{person}{Wen Jiang},
  \bibinfo{person}{Yi Wei}, {and} \bibinfo{person}{Yu Lin}.}
  \bibinfo{year}{2021}\natexlab{}.
\newblock \showarticletitle{Multiplex bipartite network embedding using dual
  hypergraph convolutional networks}. In \bibinfo{booktitle}{\emph{WWW}}.
  \bibinfo{pages}{1649--1660}.
\newblock


\bibitem[\protect\citeauthoryear{Yu, Shen, Li, and Lerer}{Yu
  et~al\mbox{.}}{2020}]%
        {yu2020scalable}
\bibfield{author}{\bibinfo{person}{Lingfan Yu}, \bibinfo{person}{Jiajun Shen},
  \bibinfo{person}{Jinyang Li}, {and} \bibinfo{person}{Adam Lerer}.}
  \bibinfo{year}{2020}\natexlab{}.
\newblock \showarticletitle{Scalable graph neural networks for heterogeneous
  graphs}.
\newblock \bibinfo{journal}{\emph{arXiv preprint arXiv:2011.09679}}
  (\bibinfo{year}{2020}).
\newblock


\bibitem[\protect\citeauthoryear{Yun, Jeong, Kim, Kang, and Kim}{Yun
  et~al\mbox{.}}{2019}]%
        {nips19GTN}
\bibfield{author}{\bibinfo{person}{Seongjun Yun}, \bibinfo{person}{Minbyul
  Jeong}, \bibinfo{person}{Raehyun Kim}, \bibinfo{person}{Jaewoo Kang}, {and}
  \bibinfo{person}{Hyunwoo~J. Kim}.} \bibinfo{year}{2019}\natexlab{}.
\newblock \showarticletitle{Graph Transformer Networks}. In
  \bibinfo{booktitle}{\emph{NeurIPS}}. \bibinfo{pages}{11960--11970}.
\newblock


\bibitem[\protect\citeauthoryear{Zhang, Song, Huang, Swami, and Chawla}{Zhang
  et~al\mbox{.}}{2019}]%
        {zhang2019heterogeneous}
\bibfield{author}{\bibinfo{person}{Chuxu Zhang}, \bibinfo{person}{Dongjin
  Song}, \bibinfo{person}{Chao Huang}, \bibinfo{person}{Ananthram Swami}, {and}
  \bibinfo{person}{Nitesh~V Chawla}.} \bibinfo{year}{2019}\natexlab{}.
\newblock \showarticletitle{Heterogeneous graph neural network}. In
  \bibinfo{booktitle}{\emph{KDD}}. \bibinfo{pages}{793--803}.
\newblock


\bibitem[\protect\citeauthoryear{Zhang, Qiu, Yi, and Song}{Zhang
  et~al\mbox{.}}{2018b}]%
        {zhang2018scalable}
\bibfield{author}{\bibinfo{person}{Hongming Zhang}, \bibinfo{person}{Liwei
  Qiu}, \bibinfo{person}{Lingling Yi}, {and} \bibinfo{person}{Yangqiu Song}.}
  \bibinfo{year}{2018}\natexlab{b}.
\newblock \showarticletitle{Scalable multiplex network embedding}. In
  \bibinfo{booktitle}{\emph{IJCAI}}, Vol.~\bibinfo{volume}{18}.
  \bibinfo{pages}{3082--3088}.
\newblock


\bibitem[\protect\citeauthoryear{Zhang, Cui, Li, Wang, and Zhu}{Zhang
  et~al\mbox{.}}{2018a}]%
        {icdm18billionNE}
\bibfield{author}{\bibinfo{person}{Ziwei Zhang}, \bibinfo{person}{Peng Cui},
  \bibinfo{person}{Haoyang Li}, \bibinfo{person}{Xiao Wang}, {and}
  \bibinfo{person}{Wenwu Zhu}.} \bibinfo{year}{2018}\natexlab{a}.
\newblock \showarticletitle{Billion-Scale Network Embedding with Iterative
  Random Projection}. In \bibinfo{booktitle}{\emph{ICDM}}.
  \bibinfo{pages}{787--796}.
\newblock


\bibitem[\protect\citeauthoryear{Zhao, Wang, Shi, Hu, Song, and Ye}{Zhao
  et~al\mbox{.}}{2021}]%
        {zhao2021heterogeneous}
\bibfield{author}{\bibinfo{person}{Jianan Zhao}, \bibinfo{person}{Xiao Wang},
  \bibinfo{person}{Chuan Shi}, \bibinfo{person}{Binbin Hu},
  \bibinfo{person}{Guojie Song}, {and} \bibinfo{person}{Yanfang Ye}.}
  \bibinfo{year}{2021}\natexlab{}.
\newblock \showarticletitle{Heterogeneous Graph Structure Learning for Graph
  Neural Networks}. In \bibinfo{booktitle}{\emph{AAAI}}.
\newblock


\end{thebibliography}


\clearpage
\appendix

\section{Supplemental Material}


\subsection{Notations}

We summarize the key notations used in the paper as well as their definitions in Table~\ref{table_notations}.

\begin{table}[htbp]
\centering
\caption{Summary of key notations}
\label{table_notations}
\begin{tabular}{c|c}
     \hline
     Notation & Definition \\
     \hline
     $\mathcal{G}$			& The target graph\\
     $\mathcal{V}, \mathcal{E}$ & the set of nodes and edges in $\mathcal{G}$\\
     $\mathcal{O},\mathcal{R}$	& the set of node and edge types in $\mathcal{G}$\\
     $\mathbf{X}$		& the matrix of node attributes in $\mathcal{G}$\\
     $\mathcal{G}_r$		& the sub-network \wrt. edge type $r$\\ 
     $\mathbf{A}_r$		& the adjacency matrix of $\mathcal{G}_r$\\
     $\mathbb{A}$       & the aggregated adjacency matrix\\
     $\mathbf{H}$       & the node embeddings\\
     $\mathbf{H}^{(l)}$       & the hidden representation for the $l$-th layer\\
     $d$			  	& the hidden dimensionality of embeddings\\
     $n, m$			  	& the number of nodes and attributes\\
     $\beta_r$                & the learnable weight for edge type $r$\\
     $\mathbf{W}^{(l)}$			    & the learnable weight matrix for the $l$-th layer\\
     \hline
\end{tabular}
\end{table}

\subsection{Algorithm Pseudo-Code}

Algorithm~\ref{alg.FastLANE} shows the pseudo-code of our proposed \system framework guided by the above objective functions (\ie, Eq.~\eqref{Unsupervised Loss} or Eq.~\eqref{Supervised Loss}). 

\begin{algorithm}[h]
\renewcommand{\algorithmicrequire}{\textbf{Input:}}
\renewcommand{\algorithmicensure}{\textbf{Output:}}
\caption{The Learning Procedure of our \system Model}
\label{alg.FastLANE}
\begin{algorithmic}[1]
\REQUIRE The generated AMHEN $\mathcal{G}$ and node feature matrix $\mathbf{X}$.
\ENSURE The node embeddings $\mathbf{H}$ of graph $\mathcal{G}$.
\STATE We generate the adjacency matrices $\{\mathbf{A}_r|r=1,2,\dots,|\mathcal{R}|\}$ by decoupling the attributed multiplex heterogeneous network into homogeneous and bipartite graphs.
\STATE Calculate $\mathbb{A} = \sum_{r=1}^{|\mathcal{R}|} \beta_r \mathbf{A}_r$
\FOR{ $i=1$ to $l$}
\STATE Calculate $\mathbf{H}^{(i)} \gets \mathbb{A}\cdot \mathbf{H}^{(i-1)} \cdot \mathbf{W}^{(i)}$
\ENDFOR
\STATE $\mathbf{H} = \frac{1}{l}(\mathbf{H}^{(1)} + \dots + \mathbf{H}^{(l)}) $
\STATE Calculate $\mathcal{L}$ using Eq.~\eqref{Unsupervised Loss} or Eq.~\eqref{Supervised Loss};
\STATE Back propagation and update parameters in \system
\STATE Return $\mathbf{H}$
\end{algorithmic}
\end{algorithm}

\subsection{Detailed Dataset Description}
i) For Alibaba dataset, four types of user-item interactions are regarded as the node-wise multiplex relationships. The item categories are considered as the ground truth labels for node classification. ii) For Amazon dataset, the multiplex edges are represented as the co-viewing and co-purchasing relations between different products. The node attributes include the external features of products, \eg, category, sales-rank, brand and price information. iii) For the AMiner dataset, three types of nodes (\ie, author, paper and conference) are included in the heterogeneous graph. The node labels are the paper domains. iv) For the IMDB dataset, movie, director and actor construct the heterogeneous nodes. We consider the genres of movies as the node labels. The bag-of-words representations are considered as the node attributed feature vectors.

v) For the DBLP dataset, four types of nodes are involved in the heterogeneous graph, namely, author, paper, venue, and term. We regard the research field of authors as the node class labels.






\subsection{Detailed Experimental Settings}

For link prediction task, we treat the connected nodes in network as positive node pairs, and consider all unlinked nodes as negative node pairs. For each edge type, we divide the positive node pairs into training set, verification set and test set according to the proportion of 85\%, 5\% and 10\%. At the same time, we randomly select the same number of negative node pairs to add into training set, validation set and test set. Notice that we predict each type of edge using all types of edges in datasets, and finally take the average of all edges as the final result. In particular, the training, validation and test sets are generated with the ratio of 80\%, 10\% and 10\%, respectively. In experiments, we train a logistic regression classifier for node classification. Notice that we repeat each experiment 10 times to report average results.




For fair comparison, we uniformly set the number of training rounds to 500 for link prediction and the number of training rounds to 200 for node classification. 
For node2vec method, the parameters $p$ and $q$ for random walk control are set as 2 and 0.5, respectively. For GATNE approach, the parameters $\alpha_r$ and $\beta_r$ are set as 1 for each edge type. For the MNE, we set the dimension of additional vectors to 10, set the length of walk as 10, set the number of walks as 20. For compared neural network-based models, the learning rate is searched from the range of $\{0.0005, 0.0001, 0.005, 0.001, 0.05, 0.01,\}$. We configure the multi-head attention with 8 head-specific representation spaces, and apply the dropout ratio of 0.6. For GTN method, the number of graph Transformer layers is set as 3. For DMGI baseline, the parameters $\alpha, \beta, \gamma$ are chosen from the value range of $\{0.0001, 0.001, 0.01, 0.1\}$. The weight $w$ for self-connection is set as 3.






For FAME, we use optuna\footnote{\url{https://github.com/pfnet/optuna}} to tune the parameters over $\alpha_1,\dots,\alpha_K$ and $\beta_1,\dots,\beta_{|\mathcal{R}|}$ as described in the original paper. 
For AM-GCN, we tune loss aggregation parameters $\beta, \gamma$ in $\{0.0001, 0.001, 0.01, 0.1\}$.
For MAGNN, we set the number of independent attention mechanisms $k=4$.
For HPN, we set iterations in semantic propagation $k=3$ and value of restart probability $\alpha=0.1$. 
For HGSL, we set the number of GNN layers to 2 and the hidden layer output dimension to 64. 
For R-GCN, we set the batch size to 126, the number of GNN layers to 2, and the hidden layer dimension to 64.
For NARS, we set the number of hops to 2, and the number of feed-forward layers to 2.
For DualHGNN, we use the asymmetric operator and set $\lambda$ as 0.5.

For our \system, we set the number of convolution layers $l$ to 2, learning rate to 0.05, dropout to 0.5, and weight-decay to 0.0005. 


The configurations of system platform for efficiency evaluation are as followed. CPU: Intel Xeon E5-2660 (2.2GHz), Memory: 80GB, 2 GPU units: GeForce RTX 2080 (8G).  

The source code of our model implementation is available at \url{https://github.com/NSSSJSS/MHGCN}.

\subsection{Baselines}

The publicly source codes of baselines can be available at the following URLs: 
\begin{itemize}
    \item \textbf{node2vec} -- \url{https://github.com/aditya-grover/node2vec}
    \item \textbf{RandNE} -- \url{https://github.com/ZW-ZHANG/RandNE} 
    \item \textbf{FastRP} -- \url{https://github.com/GTmac/FastRP} 
    \item \textbf{SGC} -- \url{https://github.com/Tiiiger/SGC}
    \item \textbf{AM-GCN} -- \url{https://github.com/zhumeiqiBUPT/AM-GCN}
    \item \textbf{R-GCN} -- \url{https://github.com/BUPT-GAMMA/OpenHGNN} 
    \item \textbf{HAN} -- \url{https://github.com/Jhy1993/HAN} 
    \item \textbf{NARS} -- \url{https://github.com/BUPT-GAMMA/OpenHGNN} 
    \item \textbf{MAGNN} -- \url{https://github.com/cynricfu/MAGNN}
    \item \textbf{HPN} -- \url{https://github.com/BUPT-GAMMA/OpenHGNN}
    \item \textbf{PMNE} -- The source code of PMNE used in this work is released by the authors of MNE at  \url{https://github.com/HKUST-KnowComp/MNE}
    \item \textbf{MNE} -- \url{https://github.com/HKUST-KnowComp/MNE}
    \item \textbf{GATNE} -- \url{https://github.com/THUDM/GATNE}
    \item \textbf{GTN} --  \url{https://github.com/seongjunyun/Graph_Transformer_Networks}
    \item \textbf{DMGI} -- \url{https://github.com/pcy1302/DMGI}
    \item \textbf{FAME} -- \url{https://github.com/ZhijunLiu95/FAME}
    \item \textbf{HGSL} -- \url{https://github.com/Andy-Border/HGSL}
    \item \textbf{DualHGNN} -- \url{https://github.com/xuehansheng/DualHGCN}
\end{itemize}

For homogeneous network embedding methods and heterogeneous network embedding methods to deal with multiplex networks, we feed separate graphs with a single-layer view into them to obtain different node embeddings, then perform mean pooling to generate final node embedding. 
Since DualHGNN is designed only for multiplex bipartite networks, it can only work on Alibaba network. 

The network types handled by the baseline methods are summarized in Table~\ref{tab:method}. 

\begin{table}[t]
\begin{center}
\vspace{-0mm}
\caption{The types of graphs handled by different methods (Heter.: Heterogenous node and edge types, Multi.: Multiplex edges, Attr.: Node attributed information, Unsup.: Unsupervised learning, Auto.: Automatic meta-path).}
\label{tab:method}
\vspace{-0mm}
\setlength{\tabcolsep}{2.0mm}{}
\begin{tabular}{c|cccccc}
\toprule
\multirow{2}{*}{Method} & \multicolumn{2}{c}{Heter.} & \multirow{2}{*}{Multi.} & \multirow{2}{*}{Attr.} & \multirow{2}{*}{Unsup.} & \multirow{2}{*}{Auto.} \\
 & Node & Edge &  &  &  &  \\
\midrule
node2vec & $\times$ & $\times$ & $\times$ & $\times$ & \checkmark & $\times$ \\
RandNE & $\times$ & $\times$ & $\times$ & $\times$ & \checkmark & $\times$ \\
FastRP & $\times$ & $\times$ & $\times$ & $\times$ & \checkmark & $\times$ \\
SGC & $\times$ & $\times$ & $\times$ & \checkmark & \checkmark/$\times$ & $\times$ \\
AM-GCN & $\times$ & $\times$ &  $\times$ & \checkmark & $\times$ & $\times$ \\
R-GCN & \checkmark & \checkmark &  $\times$ &  \checkmark & \checkmark/$\times$ & $\times$ \\
HAN &   \checkmark & \checkmark &  $\times$ & \checkmark & $\times$ & $\times$ \\
NARS & \checkmark & \checkmark &  $\times$ &  \checkmark & $\times$ & $\times$ \\
MAGNN & \checkmark & \checkmark &  $\times$ & \checkmark & \checkmark/$\times$ & $\times$ \\
HPN & \checkmark & \checkmark &  $\times$ & \checkmark & \checkmark/$\times$ & $\times$ \\
PMNE & $\times$ & \checkmark &  \checkmark & $\times$ & \checkmark & $\times$ \\
MNE & $\times$ & \checkmark &  \checkmark & $\times$ & \checkmark & $\times$ \\
GATNE & \checkmark & \checkmark &  \checkmark & \checkmark & \checkmark & $\times$ \\
GTN & \checkmark & \checkmark &  \checkmark &  \checkmark & $\times$ & \checkmark \\ 
DMGI &  \checkmark & \checkmark &  \checkmark & \checkmark & \checkmark & $\times$ \\
FAME & \checkmark & \checkmark &  \checkmark & \checkmark & \checkmark & \checkmark \\ 
HGSL & \checkmark & \checkmark &  \checkmark & \checkmark & $\times$ & $\times$ \\ 
DualHGNN & \checkmark & $\times$ &  \checkmark & \checkmark & \checkmark & $\times$ \\\hline
\model & \checkmark & \checkmark &  \checkmark & \checkmark & \checkmark/$\times$ & \checkmark \\
\bottomrule
\end{tabular}    
\end{center} 
\vspace{-0mm}
\end{table}

\subsection{Additional Experimental Results}

\subsubsection{Model Efficiency Analysis}

We also compare the efficiency of our \system with other GNN baselines for semi-supervised node classification. We report the experimental results on four datasets in Table~\ref{time}. 

As can be seen from Table~\ref{time}, 
our \system achieves the fourth-best performance after three heterogeneous network embedding methods (\ie, R-GCN, NARS and HPN). However, from the above experimental results (Tables~\ref{table_linkprediction} and \ref{tab:node_calssfication}), \system is significantly better than these three methods in both link prediction and node classification. 
\system is significantly faster than the best performed GNN baseline in node classification task (\ie, HGSL) on all datasets under the same number of training rounds. 
More specifically, our \system achieves up to 135$\times$ speedup over state-of-the-art embedding method HAN. 
\system is faster than state-of-the-art AMHEN embedding method GTN by 21.25 times on multiplex Alibaba network. 
\system is even 2.33 times and 16.58 times faster than state-of-the-art heterogeneous GNN model MAGNN on Alibaba and AMiner, respectively. 
The main reason is because our \system adopts the idea of simplifying graph convolutional networks, that is, omitting non-linear activation function. Therefore, the training efficiency of \system can be significantly improved. 
In fact, according to the above experimental results in Figure~\ref{fig:round-Ma-F1}, our model can converge quickly within 80 rounds for node classification on four tested datasets, that is, 
our model does not need to be trained for 200 rounds set in our experimental evaluation and thus can achieve faster efficiency.

\begin{table}[h]
\begin{center}
\caption{Runtime comparison of GNN methods (Second)}
\label{time}
\begin{threeparttable}
\setlength{\tabcolsep}{2mm}{}
\begin{tabular}{c|c|c|c|c}
\toprule
Method & AMiner & Alibaba & IMDB & DBLP \\
\midrule

AM-GCN &  8703.71  & 2519.82  & 24280.12  & 2786.73 \\
R-GCN & \textbf{153.04} & 301.25 & 155.40 & 192.85\\
HAN & 87105.55  & 4226.95 & 70510 & 22315.36 \\
NARS & 172.21 & \textbf{211.54}  & \textbf{75.81} & \textbf{108.54} \\
MAGNN & 10361.20  & 2320.62   & 731.03 & 2125.33 \\
HPN & 172.82 & 249.47  & 176.64 & 109.49 \\ 
GTN & OOM  & 21166.83  & 4287.20 & 18233.64 \\ 
HGSL & 1684.03 & 2120.93 & 1758.21 & 2037.10 \\ 
DualHGN & / & 11295.92 & / & /\\
\hline
\textbf{\system}& 645.20 & 996.52  & 677.23 & 970.29\\
\hline
Speedup\tnote{*}& \textbf{135.05$\times$} & \textbf{4.37$\times$}  & \textbf{104.15$\times$} & \textbf{23.01$\times$} \\
\hline
Speedup\tnote{**}& / & \textbf{21.25$\times$}  & \textbf{6.33$\times$} & \textbf{18.80$\times$} \\
\bottomrule
\end{tabular}
\begin{tablenotes}
\item [*]Speedup of \system over HAN. 
\item [**]Speedup of \system over GTN. 
\item OOM: Out Of Memory. 
\end{tablenotes}
\end{threeparttable}    
\end{center}
\end{table}



\end{document}